\documentclass[fleqn,usenatbib]{mnras}

\usepackage{amsmath,amssymb}
\usepackage[T1]{fontenc}

\usepackage{graphicx}	
\usepackage{amsmath}	
\usepackage{multicol}
\usepackage{multirow}
\usepackage{tablefootnote}
\usepackage{verbatim}
\usepackage{array}
\usepackage{soul}
\usepackage{xcolor}
\newcolumntype{P}[1]{>{\centering\arraybackslash}p{#1}}
\usepackage{ulem}

\title[Environment of dwarf AGN]{The environment of AGN dwarf galaxies at z$\sim$0.7 from the VIPERS survey}

\author[M. Siudek, et al.]{
M. Siudek,$^{1,2}$\thanks{E-mail: msiudek@ifae.es}, 
M. Mezcua,$^{2,3}$,
J. Krywult$^{4}$
\\
$^{1}$Institut de Física d’Altes Energies (IFAE), The Barcelona Institute of Science and Technology, 08193 Bellaterra (Barcelona), Spain\\
$^{2}$ Institute of Space Sciences (ICE, CSIC), Campus UAB, Carrer de Magrans, 08193 Barcelona, Spain\\
$^{3}$ Institut d'Estudis Espacials de Catalunya (IEEC), Carrer Gran Capit\`a, 08034 Barcelona, Spain\\
$^{4}$ Institute of Physics, Jan Kochanowski University, ul. Uniwersytecka 7, 25-406 Kielce, Poland
}

\date{Accepted XXX. Received YYY; in original form ZZZ}

\pubyear{2022}

\begin{document}
\label{firstpage}
\pagerange{\pageref{firstpage}--\pageref{lastpage}}
\maketitle

\begin{abstract}
Dwarf galaxies are ideal laboratories to study the relationship between the environment and AGN activity. 
However, the type of environments in which dwarf galaxies hosting AGN reside is still unclear and limited to low-redshift studies ($\rm{z<0.5}$). 
We use the VIMOS Public Extragalactic Redshift Survey (VIPERS) to investigate, for the first time, their environments at $0.5<\rm{z}<0.9$. 
We select a sample of 12,942 low-mass ($\rm{log}(M_\mathrm{*}/M_{\odot})\leq10$) galaxies 
and use the emission-line diagnostic diagram to identify AGN. 
We characterise their
local environments as the galaxy density contrast, $\delta$, derived from the fifth nearest neighbour method. 
Our work demonstrates that AGN and non-AGN dwarf galaxies reside in similar environments at intermediate redshift 
suggesting that the environment is not an important factor in triggering AGN activity already since $\rm{z=0.9}$.  
Dwarf galaxies show a strong preference for low-density environments, independently of whether they host an AGN or not. 
Their properties 
do not change when moving to denser environments, suggesting that dwarf galaxies are not gas-enriched due to environmental effects. 
Moreover, AGN presence does not alter host properties supporting the scenario that AGN feedback does not impact the star formation of the host. 
Lastly, AGN are found to host over-massive black holes. 
This is the first study of dwarf galaxies hosting AGN at $\rm{z>0.5}$. The next generation of deep surveys will reveal whether or not such lack of environmental trends is common also for faint higher-redshift dwarf galaxy populations. 
\end{abstract}



\section{Introduction}\label{sec:introduction}
Environment is found to play a significant role in shaping the star formation rate (SFR) of both low-mass ($\rm{log}(M_\mathrm{*}/M_{\odot})\leq10$) and massive ($\rm{log}(M_\mathrm{*}/M_{\odot})>10$) galaxies at z $\leq$ 1 (e.g. \citealt{Lewis2002}; \citealt{Gomez2003}; \citealt{Peng2010,Peng2012}; \citealt{Geha2012}; \citealt{Stierwalt2015}; \citealt{Darvish2016}; \citealt{Kawinwanichakij2017}; \citealt{Siudek2022}), with star-forming galaxies being more commonly found in low-density (LD) regions with a large supply of cold gas. The presence of an active galactic nucleus (AGN) has also been found by several studies to be strongly dependent on environment in the case of massive galaxies, with higher AGN fractions being found in field and underdense regions (e.g. \citealt{Kauffmann2004}; \citealt{Silverman2009}; \citealt{Sabater2013}; \citealt{Ehlert2014}; \citealt{Lopes2017}; although no to very weak environmental dependence is found by other works, e.g. \citealt{Martini2007}; \citealt{Pimbblet2013}; \citealt{Amiri2019}; \citealt{Man2019}).

Investigating environmental effects on triggering AGN activity is particularly important in dwarf galaxies, as they are the most abundant galaxies in the Universe and the building blocks of massive galaxies. 
Dwarf galaxies are in addition expected to host the seed intermediate-mass black holes (with black hole masses M$_\mathrm{BH}=100-10^{6}$ M$_{\odot}$) from which supermassive black holes grow (see \citealt{Mezcua2017}; \citealt{Greene2020}; \citealt{Reines2022} for reviews). Observational evidence for the presence of intermediate-mass black holes in dwarf galaxies comes from dynamical mass measurements (limited to the Local Group; e.g. \citealt{Nguyen2017,Nguyen2018,Nguyen2019}) and from the detection of low-mass (M$_\mathrm{BH} \lesssim 10^{6}$ M$_{\odot}$) AGN in dwarf galaxies in the local Universe (e.g. \citealt{Reines2013}; \citealt{Baldassare2015,Baldassare2017}; \citealt{Chilingarian2018}; \citealt{Woo2019}; \citealt{Mezcua2020}; \citealt{Polimera2022}; \citealt{Salehirad2022}) and out to z$\sim$ 3.4 (e.g. \citealt{Mezcua2016,Mezcua2018b,Mezcuaetal2019}). Whether such AGN activity could be triggered by galaxy mergers or environmental effects is however far from clear. 

On one hand, both cosmological simulations (e.g. \citealt{Fakhouri2010}; \citealt{Deason2014}) and large surveys (e.g. \citealt{Stierwalt2015}; \citealt{Paudel2018}) show that dwarf galaxy mergers can be very frequent, and these can ignite or enhance AGN activity. Indeed, numerous AGN are being found in dwarf galaxies undergoing a minor merger event (e.g. \citealt{Bianchi2013}; \citealt{Secrest2017}; \citealt{Mezcua2015,Mezcua2018b}; \citealt{Barrows2019}; \citealt{Kim2020}) or having a companion (\citealt{Mezcua2021}). On the other hand, most of the dwarf galaxies hosting AGN are found to be of late-type or star-forming (e.g. \citealt{Jiang2011}; \citealt{Mezcua2016,Mezcua2018a}; \citealt{Mezcua2020}; \citealt{Kimbrell2021}) and thus expected to reside in cold-gas-rich field environments (e.g. \citealt{Geha2012}; \citealt{Sabater2013}). 
Yet, no significant dependence of AGN activity with galaxy interactions nor environment has been found for dwarf galaxies at z$\leq0.055$ (\citealt{Bradford2018}; \citealt{Manzano-King2020}; \citealt{Kristensen2020}) and out to $\rm{z<0.5}$ (\citealt{Davis2022}).  
In particular, \cite{Kristensen2021} find that non-AGN dwarf galaxies in the Illustris simulation tend to have been residing in dense environments for long times, while galaxies with intermediate AGN activity have commonly been in a recent ($\leq$ 4 Gyr) minor merger. 
The recent finding of AGN in isolated and quiescent dwarf galaxies seems instead to indicate that AGN feedback rather than a dense environment is responsible for quenching star formation in these galaxies (e.g. \citealt{Bradford2018}; \citealt{Dickey2019}). Distinguishing between these two processes is crucial not only for studies of galaxy formation and evolution but also for understanding whether AGN in today's dwarf galaxies are the relics of the early Universe seed black holes (\citealt{Mezcua2019a}).  

In this paper, we investigate the dependence of AGN activity on local environment in low-mass galaxies using the VIMOS Public Extragalactic Redshift Survey~\citep[VIPERS,][]{scodeggio2018}.  Thanks to its multi-wavelength coverage and a wealth of auxiliary data, the VIPERS dataset is particularly suitable for studies of the relation of galaxy properties with the local densities. It allows us to extend the redshift of previous studies to $\rm{z=0.9}$, to investigate any environmental redshift evolution, and to circumvent previous biases caused by the use of SDSS fiber spectroscopy and emission-line-based only AGN selection techniques (e.g. \citealt{Kristensen2020}).

The paper is organized in the following way: 
in Sect. \ref{sec:data} we present the VIPERS data sample, in 
Sec.~\ref{sec:results} we show the results of environmental dependence of dwarf galaxies hosting or non-hosting AGN, and in Sec.~\ref{sec:discussion} we discuss their physical interpretation. 
Summary and conclusions are provided in Sect.~\ref{sec:summary}. 
Throughout this paper a cosmological framework with $\Omega_{m}$ = 0.3, $\Omega_{\Lambda}$ = 0.7, and $H_{0}=70$ $\rm{km s^{-1} Mpc^{-1}}$ is assumed.

\section{Data and analysis}\label{sec:data}

The sample used in this paper is retrieved from the VIPERS, completed spectroscopy survey carried out at the European South Observatory (ESO) Very Large Telescope with the VIMOS spectrograph~\citep[VIMOS,][]{lefevre03}. 
VIPERS mapped 86,775 galaxies at a redshift range of $\rm{0.5<z<1.2}$ to a limited magnitude of $\rm{i_{AB}\leq22.5}$ mag covering an area of $\sim$23.5~$\deg^2$  within the W1 (15.7~$\deg^2$) and W4 (7.8~$\deg^2$) fields of the Canada-France-Hawaii Telescope Legacy Survey Wide (CFHTLS-Wide). 
The low-resolution grid provides a spectral resolution of $\rm{R\sim220}$ with a wavelength coverage of $5,500-9,500$ \r{A}. 
Observations were carried out with a slit width of 1 arcsec and a dispersion of 7.14 \r{A} per pixel. 
The redshift uncertainty ($\rm{\sigma_{z}=0.00054\times(1+z)}$) scales  with redshift, but the linear assumption is not necessarily entirely correct as the resolution and sensitivity of VIMOS spectra are not constant and independent on the wavelength. 
The spectra signal-to-noise ratios (SNR) are  mostly driven by the galaxy apparent magnitudes but are also influenced by the observing conditions. 
The selection in spectra SNR is reflected by redshift flag ($\rm{z_{flag}}$), with high confidence redshift ($>90\%$) estimations reserved for spectra with high SNR ($\rm{z_{flag}=4}$) and clear spectral features ($\rm{z_{flag}=3}$;  $\rm{z_{flag}=9}$ for one single clear spectral emission feature), while low confidence redshifts ($\lesssim50\%$) are based on weak spectral features and/or continuum shape ($\rm{z_{flag}=1}$). 
In all VIPERS papers, galaxies with a high-confidence redshift  (i.e with $\rm{z_{flag}}$ between 2 and 9) are referred to as reliable (secure) redshift and are the only ones used in the science analyses.   
A~detailed description of the survey can be found in~\cite{guzzo} and \cite{scodeggio2018}.
The data reduction pipeline and redshift quality system are described by \cite{garilli14}.

\subsection{Physical properties}\label{sec:methods:SED}

Physical properties and absolute magnitudes were derived from spectral energy distribution (SED) fitting based on the $\rm{u}$, $\rm{g}$, $\rm{r}$, $\rm{i}$, $\rm{z}$ fluxes from the CFHTLS T0007 release~\citep{goranowa2009}, far- (FUV) and near-ultraviolet (NUV) measurements from GALEX~\citep{Martin2005}, near-infrared $\rm{Ks}$ band from WIRCAM~\citep{Puget2004} and $\rm{K_{video}}$ from the VIDEO VISTA survey~\citep{Jarvis2013}. 
The fitting process was performed with the grid of stellar population models of \cite{bruzual} fitted with the LePhare code~\citep{Arnouts2002,Ilbert2006}. 
Models are generated with the assumption of a declined star formation history with sub- and solar metallicities and the initial mass function (IMF) given by \cite{chabrier}. 
Three extinction laws (\citealp{prevot,calzetti}, and an intermediate-extinction curve from~\citealp{arnouts}) and a dust reddening of $\rm{E(B-V) \leq 0.5}$ are assumed. The emission-line contribution was taken into account following the relation between UV and line fluxes given by~\cite{Ilbert2009}. 
Stellar masses ($\rm{M_{*}}$) correspond to the median of the $\rm{M_{*}}$ probability distribution marginalised over all other fitted parameters following~\cite{ilbert2013}. 
We compare the $\rm{M_{*}}$ estimations with the ones derived for the VIPERS AGN Type II sample of~\cite{Vietri2022}. These authors used another SED fitting tool~\citep[CIGALE;][]{Boquien2019} as well as other prescriptions to generate a grid of templates and included the AGN models of~\cite{Fritz2006}. The median difference between the $\rm{M_{*}}$ used in this paper and the ones in~\cite{Vietri2022} is smaller than the error of the $\rm{M_{*}}$ estimations, indicating that the $\rm{M_{*}}$ are robust and independent on the SED fitting methodology.  
We note that the lack of AGN templates in the SED fitting procedure used in this paper may, at most (if at all), lead to an overestimation of the $\rm{M_{*}}$ of AGN (as the part of the fitted light does not come from the host galaxy), from which our low-mass galaxy sample would only benefit by moving to an even lower $\rm{M_{*}}$ regime. 
The detailed description of the physical parameters (absolute magnitudes, $\rm{M_{*}}$ and SFRs) for the VIPERS sample used in the following analysis and computed via SED fitting can be found in~\cite{moutard16b}.

\subsubsection{Star formation rates}\label{sec:SFR(OII)}

The SFR estimated through SED fitting (SFR(SED)) is sensitive to the wavelength coverage of the observed photometric data and the lack of far-infrared observations may alter its estimations~\citep[e.g.][]{Ciesla2015}. 
Since VIPERS observations in the far infrared are limited and not incorporated in the SED fitting (see Sec.~\ref{sec:methods:SED}), we use an alternative SFR indicator based on the [OII]$\lambda$3726 line~\citep[SFR(OII); e.g.][]{Kennicutt1998,Hopkins2003,Kewley2004}. 
The [OII] line doublet is unresolved in VIPERS spectra and we fit its profile using a single Gaussian model (see Sec.~\ref{sec:BPT} for details about line measurements). 
We correct the emission lines for extinction following~\cite{calzetti} assuming $\rm{R_v}=4.05$ and using the $\rm{E(B-V)}$ derived from the SED fitting. 
In addition, the $\rm{H\beta}$ line is corrected for stellar absorption following~\cite{Hopkins2003} and assuming a correction of the equivalent width (EW) of 2\r{A}, as commonly used in the literature~\citep[e.g.][]{Kong2002,Goto2003,Pistis2022}. 

We use the same SFR(OII) indicator as in the analysis of VIPERS AGN activity of \cite{Vietri2022}, following the definition given by~\cite{Kewley2004}:
\begin{equation}
 \rm{   SFR(OII)(M_{\odot}yr^{-1})= 6.58\pm1.65\times10^{-42}L(OII)(ergs^{-1}),}
\end{equation}
where L(OII) corresponds to the [OII]$\lambda$3726 luminosity.
As shown in Sec.~\ref{sec:MS} we find a good agreement between the SFR(SED) and the SFR(OII) estimations further strengthening the robustness of the physical properties derived through SED fitting. 

\subsection{Morphology}~\label{sec:data_morphology}
The structural parameters for VIPERS galaxies were derived with GALFIT~\citep{peng2002} by~\cite{krywult}. 
Briefly, for each detected galaxy in the i-band CFHTL-Wide image the S\'ersic model was convolved with a local point spread function (PSF).
The best fit was found by minimising the goodness, $\chi^2$, of the fit. 
GALFIT provides the semi-major axis ($a_\mathrm{e}$), the axial ratio ($b/a$) of the profile, from which the circularised effective radius ($R_\mathrm{e}=a_\mathrm{e}\sqrt{b/a}$) was derived, as well as the S\'ersic index $n$.  
Simulated galaxies using GALFIT added into the CFHTLS images were used to test the accuracy of the derived parameters, returning the errors of $R_\mathrm{e}$ measurements at the level of 4.4$\%$ (12$\%$) for 68$\%$ (95$\%$) of the VIPERS sample~\citep{krywult}. 
These tests prove that the automatic estimations of structural parameters are robust in preselecting VIPERS galaxies based on their shapes and can be used to investigate their change with the environment. 
A~detailed description of the structural parameters can be found in~\cite{krywult}.

\subsection{Environment}\label{sec:env}

We use the local density contrast, $\delta$, from \citealt{Cucciati2017} as a measure of the environment. 
In their work the $\delta$ parameter is defined  as:
\begin{equation}
\delta(RA,DEC,\rm{z})=[\rho(RA,DEC,\rm{z})-\langle\rho(\rm{z})\rangle]/\langle\rho(\rm{z})\rangle,
\end{equation}
where $\rho(RA,DEC,\rm{z})$ is the local density of the tracer (all the galaxies of the sample used to trace density field) centered at the  galaxy $(RA,DEC,\rm{z})$, and $\langle\rho(\rm{z})\rangle$ is the mean density at that redshift. 
The density field is computed using a cylinder with a half-length of  $\pm1000$ km/s and the radius equal to the distance of the fifth nearest neighbour ($\rm{D_{p,5}}$). 
Galaxies that trace the density field (tracers) were selected from a volume-limited sample that included galaxies with both spectroscopic and reliable photometric redshifts (see more details in~\citealt{Cucciati2017}). 
In this paper, we use tracers with cut $\rm{M_{B} \le (-20.4 - z)}$, which provides the completeness within redshift range of  $\rm{0.5<z\leq0.9}$. 
The resolution of the density contrast would be higher if we used fainter tracers but the redshift limit would be lower since faint galaxies can only be observed at lower redshifts. 
This tracer selection provides a comoving number density that does not evolve, therefore is not affected by discreteness effects that change with redshift. 
Moreover, this tracer cut is recommended for scientific analysis~\citep{Cucciati2017} and was widely adapted in VIPERS galaxy environmental studies~\citep[][Siudek et al. submitted]{davidzon16,Cucciati2017,Gargiulo2019,Siudek2022}.  

\citealt{Cucciati2017} defined the thresholds for LD and high-density (HD) environments based on the quartiles of the $\delta$ distribution of the full population of VIPERS galaxies (i.e. including both massive red and low-mass blue galaxies). 
The LD environment corresponds to the first percentile ($\rm{log(1+\delta)}\leq0.24$) and the HD environment to the fourth quartile ($\rm{log(1+\delta)}>0.72$) of $\delta$ distribution. 
Similar thresholds were found by other before-mentioned authors using VIPERS $\delta$ measurements, so we also adapt the same definitions in our work. 
Roughly, the LD environment catches the void galaxies with an average projected $\rm{D_{p,5}}$ $\sim3.5\rm{h^{-1}Mpc}$, while the HD environment characterises group and cluster galaxies with an average projected $\rm{D_{p,5}}$ $\sim2.0\rm{h^{-1}Mpc}$~\citep[although the 
dimension is still too high, see details in][]{Cucciati2017}.   
A~detailed description of the environment properties can be found in~\citealt{Cucciati2014, Cucciati2017}.

\subsection{Sample selection}\label{sec:sample_selection}
We select a sample of 33,333 low-mass VIPERS galaxies fulfilling the following criteria: i) having $\rm{log}(M_\mathrm{*}/M_{\odot})\leq10.0$, and ii) having a reliable redshift measurement (i.e 
with a confidence level of 90\% or larger, which are the only ones recommended for scientific analyses; \citealt{garilli14, scodeggio2018}).
The $\rm{M_{*}}$ threshold adopted here is commonly used in studies of dwarf galaxies, but is slightly higher than that of searches for AGN in dwarf galaxies (where the $\rm{M_{*}}$ of the Large Magellanic Cloud (LMC) is typically adopted; $\rm{log}(M_\mathrm{*}/M_{\odot})\sim9.5$,~\citealt{vanderMarel2002}; e.g.~\citealt{Reines2013, Mezcua2016, Mezcua2018a, Mezcua2019b,Polimera2022}). 
However, different studies also adopt slightly higher cut on $\rm{M_{*}}$ to define dwarf galaxies (e.g.~\citealt{Manzano-King2020} uses $\rm{log}(M_\mathrm{*}/M_{\odot})\leq10$ to study AGN-driven kinematics in dwarf galaxies). 
Moreover, it is expected that a high fraction of $\rm{log}(M_\mathrm{*}/M_{\odot})=9.0-10.0$ galaxies host intermediate-mass black holes~\citep{Greene2020}. 
As we intent to create a sample of dwarfs hosting AGN large enough to study environmental effects, we relax the LMC criterion and use a threshold 
as in~\cite{Manzano-King2020}. 
In Appendix~\ref{app:sanity_check} we show that restricting the $\rm{M_{*}}$ criterion by 0.5 dex does not affect our results. 

\begin{figure}
	\centerline{\includegraphics[width=0.49\textwidth]{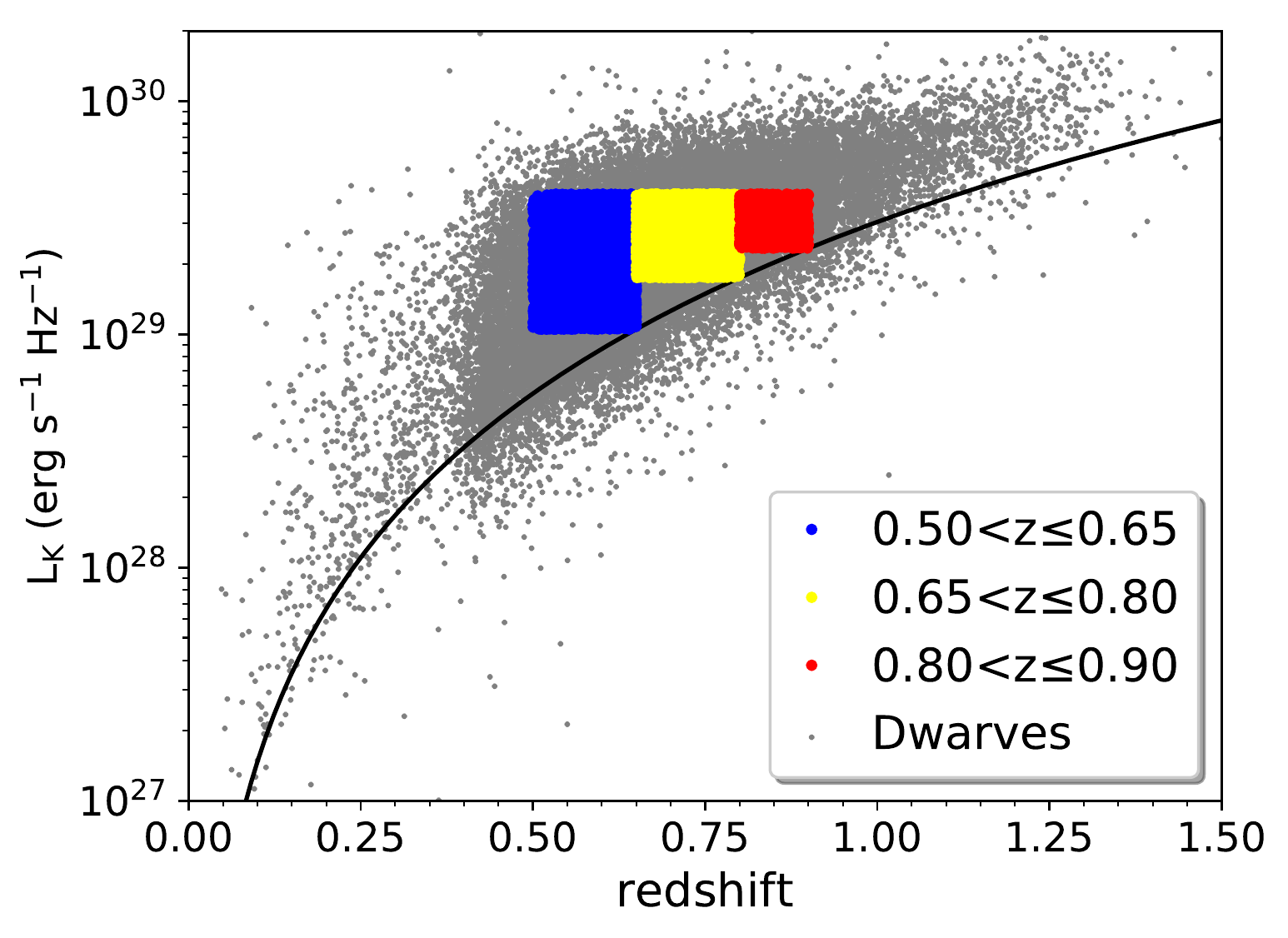}}
	\caption{K-band luminosity versus redshift for the parent sample of 33,333 VIPERS dwarf galaxies. The solid line corresponds to a Ks-band sensitivity limit of 22 mag. The three z-bins considered in this study are shown as color-coded regions.}
	\label{fig:zbins}
\end{figure}

To study any evolution of environment with redshift, we select three complete K-band luminosity ($L_\mathrm{K}$) redshift bins ($0.5<\rm{z}\leq0.65$, $0.65<\rm{z}\leq0.80$, and $0.80<\rm{z}\leq0.90$; see Fig.~\ref{fig:zbins}). 
Such redshift bins have two advantages: (i) they probe a  sufficiently large volume in each bin ($\gtrsim 7\times10^6\rm{h}^{-3}\rm{Mpc}^3$), and (ii) their median redshifts correspond to nearly equally-spaced time intervals of 0.6 - 0.7 Gyr. 
We follow the same approach as in~\cite{Mezcua2016}: each $L_\mathrm{K}-\rm{z}$ bin includes galaxies with $L_\mathrm{K}$ above that of the Ks (or $\rm{K_{video}}$ in case of lack of measurements in Ks) magnitude limit of 22~\citep{moutard16a} at that redshift and below a maximum $L_\mathrm{K}$ of 10$^{30}$ erg s$^{-1}$ Hz$^{-1}$.
This $L_\mathrm{K}$-z bins limit our sample to 15,696 galaxies. 

We further restrict our sample to only galaxies with reliable $\delta$ measurements, i.e. for which at least $60\%$ of the cylinder volume is within survey footprint (gaps and boundaries, see details in \citealt{Cucciati2017}, and \citealt{davidzon16}). 
This criterion limits our final sample to 12,942 dwarf galaxies within the redshift range $0.5<z\leq0.9$. 

\subsubsection{AGN identification: emission line diagnostic diagram}~\label{sec:BPT}

To identify the presence of AGN in the final sample of 12,942 dwarf galaxies we make use of the emission line diagnostic diagram based on the [OII]$\lambda3726$, H$\beta$, and [OIII]$\lambda5007$ emission lines proposed by~\cite{2010Lamareille}, which seems to be the most comparable approach to the Baldwin, Philips and Telervich (BPT; \citealt{Baldwin81}) diagram but at higher redshift ($\rm{z>0.45}$) where $H\alpha$ and NII (standard lines used in the BPT diagram) lie outside the spectral coverage of optical spectrographs (e.g. \citealt{2010Lamareille}; \citealt{Thomas2013}). 
The [NII]/H$\alpha$ emission line ratio of the BPT diagram might be replaced not only by [OII]/H$\beta$, but also by rest-frame U-B galaxy colour (CEx diagram; \citealt{Yan2011}), absolute magnitude in H band (\citealt{Weiner2006}), 4000 \AA~break strength (DEW diagram; \citealt{Marocco2011}) or stellar mass (MEx diagram; \citealt{Juneau2011,Juneau2014}). 
Throughout this paper, we use the diagram proposed by \cite{2010Lamareille}, but for sanity check we also consider standardly used MEx diagram to identify a sample of AGN. 
Independently of the choice of the diagram, our results do not change (see App.~\ref{app:sanity_check}). 
In diagnostic diagrams that use lines that are placed far away to each other in the spectrum the line ratio is relatively sensitive to reddening effects. To overcome this problem in our analysis we use the EW of the emission lines instead of line fluxes~\citep{2010Lamareille}.

The line measurements were obtained using the EZ software~\citep{Garilli2010}. 
The algorithm fits the continuum using a running box. 
A continuum-subtracted spectrum around the expected position of the emission line is extracted and a Gaussian is fitted to the line. 
The total flux is computed by integrating the Gaussian function as resulting from the fit, in a range of $\pm 3\sigma$. 
The EW is given by the ratio of the line flux over the continuum mean value.
In addition, the EW(H$\beta$) is corrected for the average absorption component of 2~\r{A} in EW (see Sec.~\ref{sec:SFR(OII)}). 
The error on the flux takes into account the error on the continuum (computed as the mean square root around the fitted value), the Poissonian error on line counts, and the fit residuals. The error on the EW is computed by simple error propagation.

To construct the diagnostic diagram we further limit our sample by selecting only galaxies with reliable emission line measurements. 
We select reliable sources for which the measurements of [OII]$\lambda3726$, H$\beta$, and [OIII]$\lambda5007$ satisfy the following conditions:
\begin{itemize}
    \item the distance between the expected position and the
Gaussian peak must be within 7~\r{A} ($\sim$1 pixel),
    \item the full width at half maximum (FWHM) of the line must be between 7 and 22~\r{A} (from 1 to 3 pixels),
    \item the Gaussian amplitude and the observed peak flux must differ by no more than 30\%.
\end{itemize}

\begin{figure}
	\centerline{\includegraphics[width=0.49\textwidth]{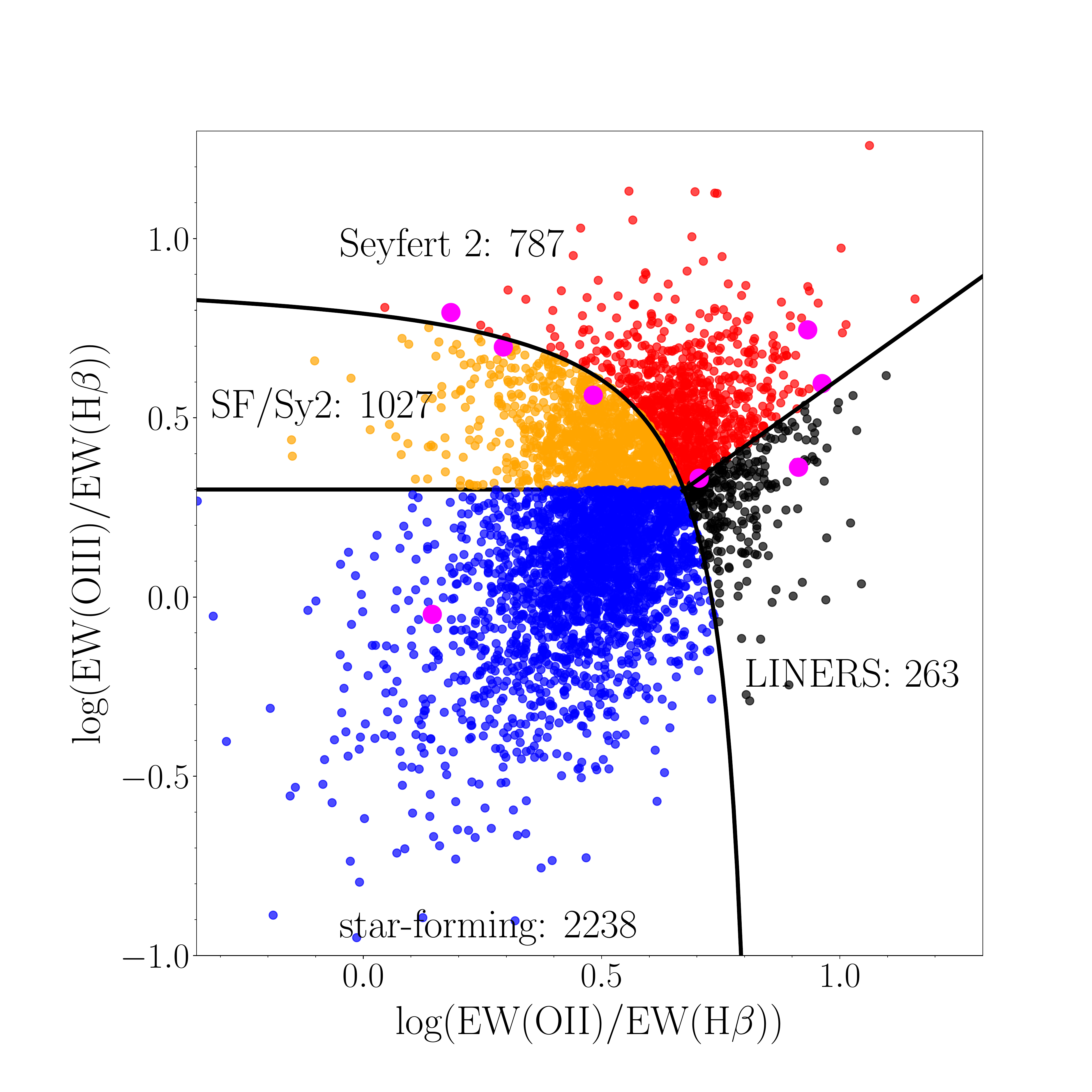}}	
	\caption{Emission line diagnostic diagram for the parent sample of 4,315 VIPERS dwarf galaxies with reliable emission line measurements. X-ray AGN are marked with pink circles.}
	\label{fig:bpt}
\end{figure}

After applying the above-mentioned criteria we end up with 4,315 dwarf galaxies.  
These conditions are recommended to create a clean sample with reliable line measurements~\citep[e.g.][]{Pistis2022,Vietri2022}. 
However, to validate if these cuts are sufficient for reliable AGN selection based on diagnostic diagrams, we also define a secure sample of 
dwarf galaxies with an additional criterion that the EW must be detected above 3$\sigma$. 
As reported in Appendix~\ref{app:sanity_check}, adopting this more restrictive criterion on the quality of emission line measurements does not change our results. 
Therefore, to preserve the large sample to study the environment properties we use a sample of 4,315 dwarf galaxies. 
The emission line diagnostic diagram for this final sample is presented in Fig.~\ref{fig:bpt}.
Based on the diagnostic diagram we distinguish a sample of 1,050 AGN composed by Seyferts and low-ionization nuclear emission-line regions (LINERs). 

\subsubsection{AGN identification: X-ray and radio counterparts}~\label{sec:xray_selection}

We also look for AGN in the sample of dwarf galaxies based on X-ray and radio diagnostics. We search for X-ray and radio counterparts within 5 arcsec of each source making use of the XMM-Newton coverage (the XMM-XXL catalogue of~\citealp{Chiappetti2018}) and the Giant Metrewave Radio Telescope (GMRT) coverage at 610 MHz (the GMRT XXL-N 610 MHz catalogue of~\citealp{Smolcic2018}), respectively, of the CFHT field. Thirteen of the 12,942 dwarf galaxies with reliable environmental measurements are found to have an X-ray detection at 2-10 keV, and eight a radio detection at 610 MHz. The k-corrected 2-10 keV luminosity of the 13 X-ray detected galaxies is L$_\mathrm{2-10 keV} \gtrsim 10^{43}$ erg s$^{-1}$ and thus fully consistent with AGN.  
The 610 MHz radio luminosity ranges from $\sim6.5 \times 10^{17}$ W Hz$^{-1}$ to $\sim1.3 \times 10^{19}$ W Hz$^{-1}$ and is thus consistent with that from X-ray binaries and radio supernovae~\citep[e.g.][]{Mezcua2013,Mezcua2019b,Reines2020}. We note that five of the 13 X-ray AGN were already classified as AGN by the emission-line diagnostic diagram (see Fig.~\ref{fig:bpt}). Hence, the final sample of AGN in dwarf galaxies consists of 1,058 sources. 
In Appendix~\ref{app:sanity_check} we verify that disregarding LINERs does not influence our results. 

	\begin{table*}
		\centering                         
		\begin{tabular}{p{1.2cm} p{1.2cm} p{1.2cm} p{1.2cm} p{1.2cm} p{1.2cm} p{1.6cm} p{1.6cm}}    
			\hline 
			sample & N & z & $\rm{M_{*}}$ & log(1+$\delta$) & SFR(SED) &  NUVr & rK \\
			\hline 
            \hline
			AGN & 1,058 & 0.64$\pm$0.09 & 9.52$\pm$0.20 & 0.23$\pm$0.37 & 0.72$\pm$0.37 & 1.62$\pm$0.41 & $0.40\pm0.29$ \\
			SF(control) & 1,058 & 0.64$\pm$0.09 & 9.53$\pm$0.20 & 0.25$\pm$0.38 & 0.75$\pm$0.38 & 1.76$\pm$0.39 & $0.43\pm0.28$ \\
			\hline 
		\end{tabular}
		\caption{Physical properties of the AGN and SF(control) samples. The number of members (N) and the mean with standard deviation for: redshift (z), stellar mass $\rm{M_{*}=log}(M_\mathrm{*}/M_{\odot})$, density log$(1+\delta)$, star formation rate $\rm{SFR=log(SFR)}$ $\rm{[M_{\odot}yr^{-1}]}$ and colours (NUVr, rK) are provided. 
			 }             
		\label{table:properties}     
	\end{table*}

\subsubsection{AGN identification: mid-infrared colours}~\label{sec:wise_selection}
\begin{figure}
	\centerline{\includegraphics[width=0.49\textwidth]{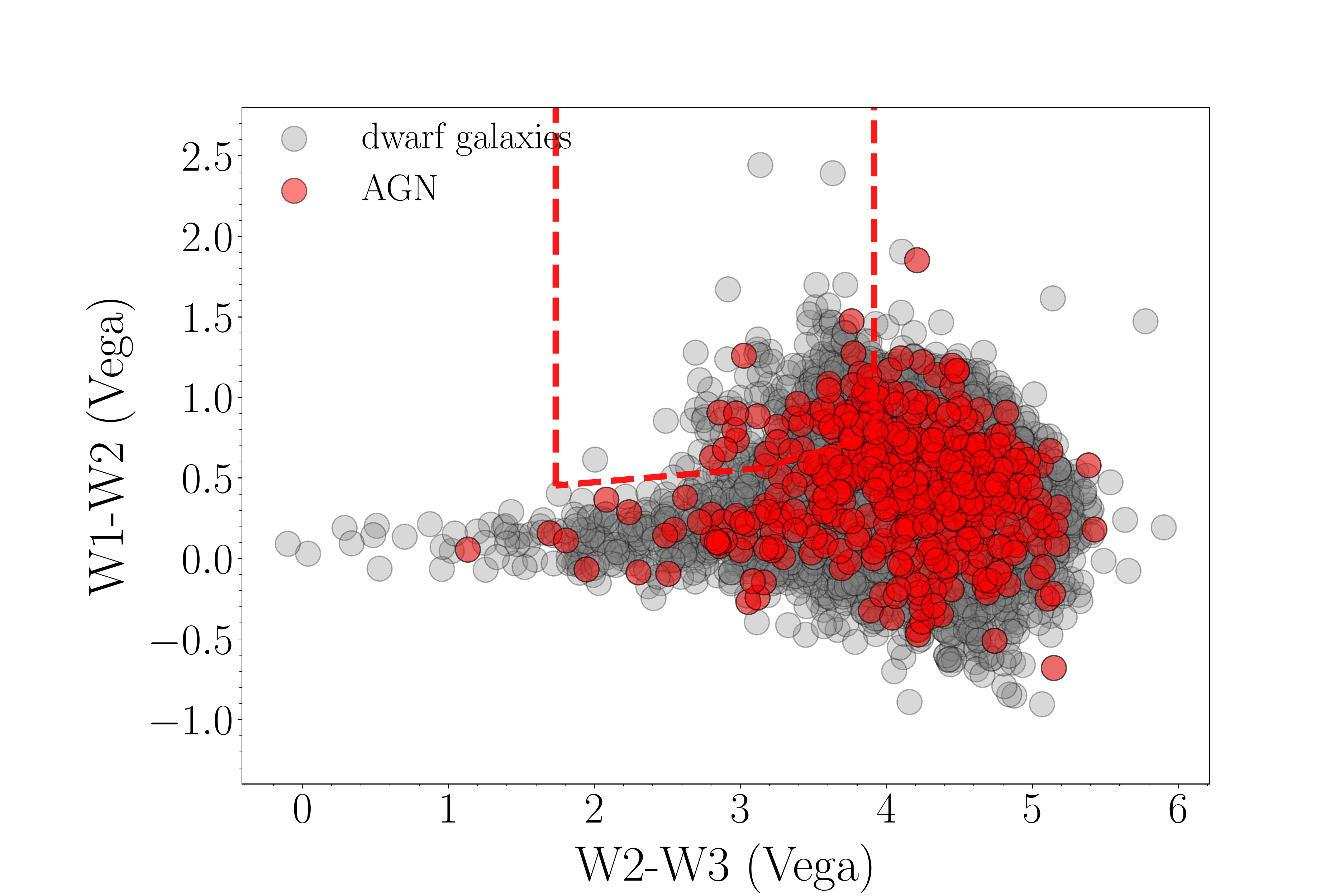}}	
	\caption{WISE colour-colour diagram for 12,942 dwarf galaxies. The red dashed line corresponds to the AGN selection proposed by~\citealp{Hviding2022}.}
	\label{fig:wise}
\end{figure}

We also identify AGN candidates based on the mid-infrared colours from the Wide-field Infrared Explorer~\citep{Wright2010AJ....140.1868W}. 
The WISE catalogue is matched to our sample of dwarf galaxies using a search radius of 10 arcsec. 
We find WISE sources observed with at least the W1, W2, and W3 filters for 5,333 out of 12,942 dwarf galaxies. 
There is a number of ways of selecting AGN using mid-infrared colours. In this work, we use a recently proposed selection criterion aimed at capturing a higher total number of spectroscopic AGN with the completeness of over 13.8\%~\citep{Hviding2022}. 
With this method, we recover $\sim10\%$ of AGN from our final AGN sample (see Fig.~\ref{fig:wise}), which is close to the completeness of the adopted cut. 
This confirms that mid-infrared selection solely is not able to recreate the full AGN spectroscopically-selected sample. 

Additionally, there are 393 WISE-selected AGN not included in our AGN sample: 277 with unreliable line measurements and 116 spread over the composite and star-forming galaxy regions on the emission line diagnostic diagram. 
We note that WISE-selected AGN candidates should be treated with a great care as star-forming dwarf galaxies that are capable to heat dust to very high temperatures can mimic the mid-infrared colours of more luminous AGN~\citep[e.g.][]{Hainline2016ApJ...832..119H}. 
Although the accuracy of the WISE selection is high ($\sim80\%$ \citealt{Hviding2022}), we prefer to preserve the purity of our sample and do not add the 393 WISE-selected AGN to our final AGN sample. In Appendix~\ref{app:sanity_check} we verify that the incorporation of these additional WISE-selected AGN would not affect our main conclusions. 

\subsubsection{Control sample}\label{sec:control_sample}

We create a control sample of star-forming galaxies (SF(control)) that matches the distributions of stellar masses, redshift, and $\rm{r-i}$ ($\rm{ri}$) colour of the AGN sample.  
Our approach to create a control sample follows similar methodology to the one proposed by~\cite{Cheung2015} and \cite{Kristensen2020}, who studied AGN and bar connection, and environmental dependence of AGN in the local Universe, respectively. 
The criteria proposed by~\cite{Cheung2015} are the following:
\begin{itemize}
    \item $\rm{|1-M_{SF}/M_{AGN}|<0.2}$,
    \item $\rm{|z_{SF}-z_{AGN}|<0.4}$,
    \item $\rm{|ri_{AGN}-ri_{SF}|<0.4}$,
\end{itemize}
where $\rm{M}$ corresponds to $\rm{log}(M_\mathrm{*}/M_{\odot})$, and subscripts $\rm{_{SF}}$, $\rm{_{AGN}}$ denote a sample of control star-forming galaxies and AGN, respectively.  
The same criteria were used by  \cite{Kristensen2020} with the following differences: i) introducing stricter redshift criterion ($|\rm{z}_{SF}-\rm{z}_{AGN}|\leq0.01$) to compensate SDSS fiber effects (from which our sample does not suffer), and ii) used $\rm{u-r}$ instead of $\rm{ri}$ colour. 
To find the SF(control) sample that is the most similar to our AGN sample we look for the closest counterpart (a nearest neighbour\footnote{\url{https://scikit-learn.org/stable/modules/neighbors.html}}) of each of our AGN dwarf galaxies in a three-dimensional space of stellar mass, redshift and ri colour within a sample of 2,237 star-forming galaxies classified based on the emission line diagnostic diagram (see Fig.~\ref{fig:bpt}). 
If the same star-forming galaxy is matched to more than one AGN we remove the duplicates and take the next best-match. 
This way we provide a unique SF(control) sample of 1,058 galaxies without double entries.
The main properties of the SF(control) sample as well as the AGN sample are given in Table~\ref{table:properties}. 
On average, $\rm{M_{*}}$, SFRs and colours of the SF(control) sample are the same as for the AGN sample (with the difference of their means $<0.2\sigma$) allowing us to narrow the before-mentioned criteria proposed by~\cite{Cheung2015} to $\rm{|1-M_{SF}/M_{AGN}|<0.1}$, $\rm{|z_{SF}-z_{AGN}|<0.2}$, and $\rm{|ri_{AGN}-ri_{SF}|\lesssim0.2}$. 
In particular, the mean difference of the main properties ($\rm{M_{*}}$, redshift and $\rm{ri}$ colour) between the AGN and SF(control) sample is $\sim0.00\pm0.04$ (see Fig.~\ref{fig:param_dif}). 
Moreover, the AGN sample has the $\rm{M_{*}}$ and ri colour indistinguishable from the SF(control) sample for all the redshift bins considered in this work as indicated by statistical tests discussed in Appendix~\ref{app:statistical_tests}. 
Furthermore, in Appendix~\ref{app:sanity_check} we verify that removing WISE-selected AGN candidates from SF(control) sample (see Sec.~\ref{sec:wise_selection}) do not change our main conclusions.

\section{Results}\label{sec:results}
In this Section, we present the environmental properties of dwarf galaxies hosting or non-hosting AGN. 
We compare the sample of emission line diagnostic diagram and X-ray selected dwarf AGN with their non-AGN counterparts (SF(control) sample).   
When computing the fractions, we weight each galaxy by a selection weight ($w$) accounting for survey incompleteness, i.e. $w=1/(TSR\times SSR\times CSR)$. 
This selection criterion includes three statistical functions: the target sampling rate ($TSR$), the spectroscopic success rate ($SSR$), and the colour sampling rate ($CSR$). 
The $TSR$ weight reflects the fraction of observed galaxies over the number of potential target galaxies. 
The $SSR$ weight represents the fraction of targeted galaxies with reliable redshift measurements and the $CSR$ weight accounts for the adopted colour cut to preselect galaxies at $\rm{z}>0.5$. 
Further details about these criteria are provided in~\cite{garilli14} and~\cite{scodeggio2018}. 
To account for the mass incompleteness introduced by the Malmquist bias each VIPERS galaxy is weighted also by the fraction of the volume in which the galaxy would be still observable (using minimal and maximal redshifts at which it could be observed), following the  $1/V_{max}$ method~\citep{Schmidt1968}. 
Summarising, the fractions are weighted by a factor $w$ that corresponds to the selection weight and $1/V_{max}$ method. 

\subsection{Density distribution}\label{sec:density_distribution}

\begin{figure*}
	\centerline{\includegraphics[width=0.99\textwidth]{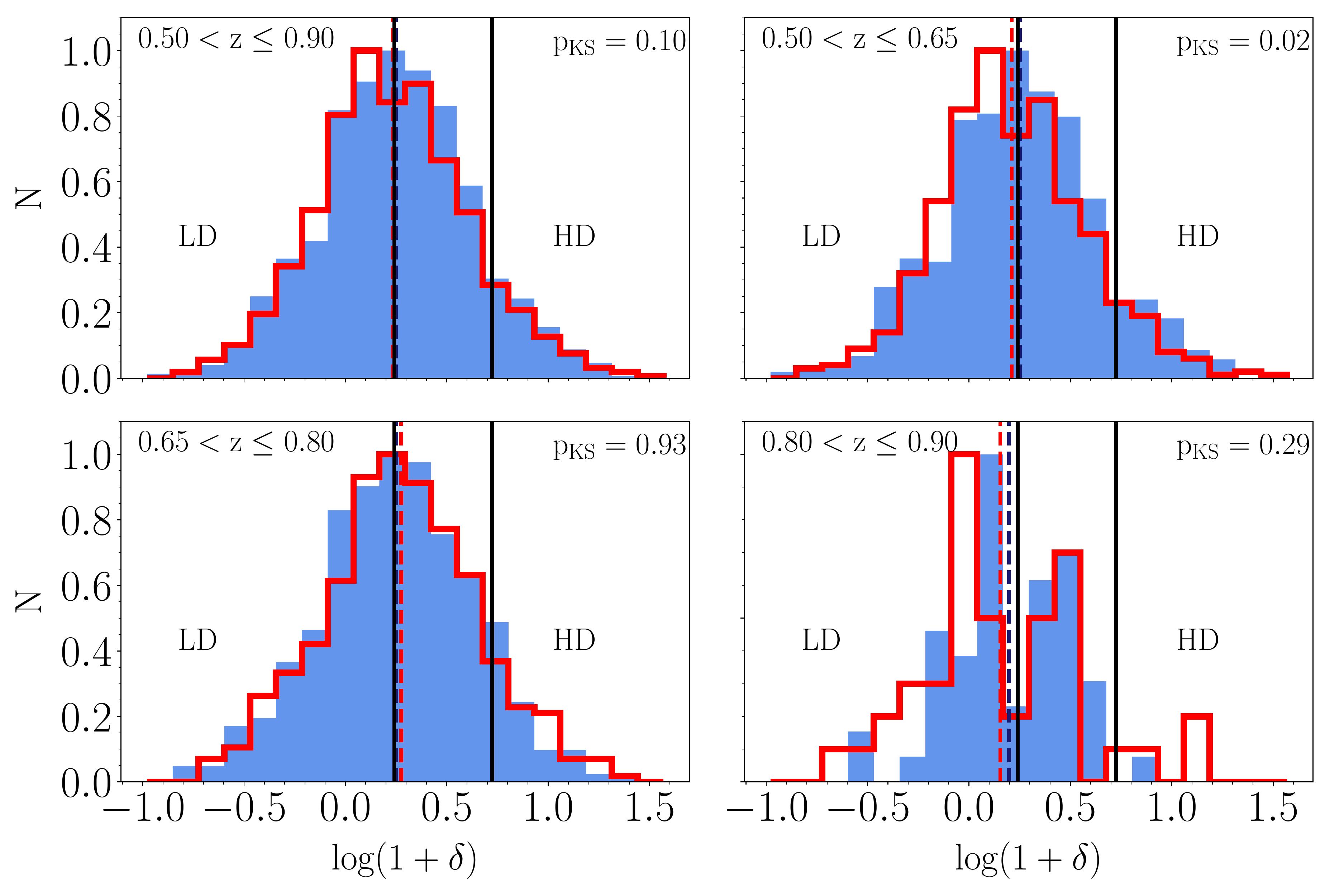}}
	\caption{In each panel the normalised distribution of the $\delta$ is given for three redshift bins and all together going clockwise. The filled blue histograms correspond to the SF(control) samples. The AGN distributions are marked in red. The first and third percentile of the VIPERS density distribution~\citep{Cucciati2017}, used to define LD and HD environments, are marked with solid black lines. The dashed lines correspond to the mean values.  
	The p-values of the KS test are displayed in the upper right of each panel.}
	\label{fig:delta_distribution}
\end{figure*} 

The $\rm{log(1+\delta)}$ distributions for the SF(control) and AGN samples over $0.5<\rm{z}\leq0.9$ are  similar as shown in the upper left panel in Fig.~\ref{fig:delta_distribution}. 
However, low p-values of the Kolmogorov-Smirnov (KS) test ($\rm{p_{KS}}=0.1$; see Appendix~\ref{app:statistical_tests} for details about statistical tests)  for AGN-SF(control) $\delta$ distributions indicate that the two distributions are probably different. 
This suggests that from a statistical point of view the $\delta$ distribution for the AGN is not the same as for the SF(control) sample. 
However, as it is clear from Fig.~\ref{fig:delta_distribution}, galaxies are preferably found in LD environments for both samples: their $\delta$ means overlap with the upper threshold for LD environment. 
As reported in Tab.~\ref{table:number_in_HD_LD}, $\sim50\%$ of the dwarf galaxies, whether  hosting or not AGN, are located in LD environments, i.e. the regions without a pervasive presence of cosmic structure. 
At the same time, only $\sim10\%$ of them are found in HD environments, i.e. groups or clusters. 
 
	\begin{table}
		\centering                         
		\begin{tabular}{p{0.09\textwidth} p{0.09\textwidth} p{0.02\textwidth} p{0.02\textwidth} p{0.02\textwidth} p{0.02\textwidth} p{0.02\textwidth}}  
		
			\hline 
			sample  & zbin & $\rm{N_{LD}}$ & $\rm{N_{HD}}$ & $\rm{\%_{LD}}$ & $\rm{\%_{HD}}$\\
			\hline 
            \hline
			AGN & 0.50$<$z$\leq$0.90 & 549 & 105 & \textbf{52} & \textbf{10}  \\
			 & 0.50$<$z$\leq$0.65 & 336 & 55 & 50 &  8 \\
			 & 0.65$<$z$\leq$0.80 & 187 & 70 & 52 &  13 \\
			 & 0.80$<$z$\leq$0.90 & 26 & 3 & 60 &  7 \\
			 \hline
			SF(control) & 0.50$<$z$\leq$0.90 & 518 & 107 & \textbf{49} & \textbf{10}   \\
			 & 0.50$<$z$\leq$0.65 & 346 & 72 & 49 &  10 \\
			 & 0.65$<$z$\leq$0.80 & 144 & 34 & 48 &  11 \\
			 & 0.80$<$z$\leq$0.90 & 28 & 1 & 54 &  2 \\
			 \hline
		\end{tabular}
		\caption{Number and percentage of galaxies for AGN and SF(control) samples found in LD and HD environments. 
			 }             
		\label{table:number_in_HD_LD}     
	\end{table}

In order to verify if the $\delta$ distribution evolves, we compare the $\delta$ distributions in three redshift bins: $\rm{0.50<z\leq0.65}$, $\rm{0.65<z\leq0.80}$, and $\rm{0.80<z\leq0.90}$, the same used to create a complete sample (see Sect.~\ref{sec:sample_selection}). 
The $\delta$ distributions in each redshift bin are presented in Fig.~\ref{fig:delta_distribution}. 
Similarly as for the $\delta$ distribution for the whole considered redshift range $\rm{0.50<z\leq0.90}$, the $\delta$ distributions for the AGN and SF(control) samples are highly probably different in the redshift bins $\rm{0.50<z\leq0.65}$ and $\rm{0.80<z\leq0.90}$ (see p-values reported in Fig.~\ref{fig:delta_distribution}). However, at redshift bin $\rm{0.65<z\leq0.80}$ there are no statistically significant differences in the distributions of AGN and SF(control) samples as indicated by high p-values ($\rm{p_{KS}=0.93}$). 
This suggests that at this redshift bin, $\rm{0.65<z\leq0.80}$ (see lower left panel in Fig.~\ref{fig:delta_distribution}), AGN and SF(control) galaxies reside in the same environments. 
Moreover, the AGN distribution at this redshift bin seems to be less asymmetric than the one in the other redshift bins.  

To verify if the asymmetry of $\delta$ distribution is statistically significant, we quantify it by measuring the skewness, $S$, the third moment of the distribution functions. 
The $\delta$ distributions for the AGN and SF(control) samples at $\rm{0.50<z\leq0.90}$ are positively skewed, however, for the AGN sample the skewness is at least three times stronger (see Tab.~\ref{tables:skewness} for the $S$ measurements). 
The skewness for $\delta$ distribution for AGN is positive, as well for the SF(control) sample at $\rm{z\leq0.65}$, but at two remaining redshift bins for the SF(control) the skewness is negative. 
The negative skewness could be explained by a galaxy bias, if galaxies preferentially form in the highest density peaks and as time passes the cluster galaxies are not anymore members of the clusters. 
On the other hand, the positive skewness points that more sources are found in the LD environments.
High positive value of the skewness, observed at the highest redshift bin, $\rm{0.80<z\leq0.90}$, for AGN sample suggests the preference of AGN towards LD environments. This is also confirmed by the higher fraction of AGN at $\rm{0.80<z\leq0.90}$ (60\%) with respect to lower bins (50\% and 53\% for $\rm{0.50<z\leq0.65}$ and $\rm{0.65<z\leq0.80}$, respectively, see Tab.~\ref{table:number_in_HD_LD}). 
We note though that the relatively small sample size of the AGN sample in the highest redshift bin makes it difficult to confirm that the positive skewness is a physical effect of the evolution. 
Interestingly, at the redshift bin $0.65<z\leq 0.80$ the skewness is close to 0, implying a negligible asymmetry in the $\delta$ distribution. 
At the same time, AGN $\delta$ distribution at this redshift bin mimic the skewness found for SF(control) sample confirming that the $\delta$ distributions are indistinguishable for AGN and SF(control) samples at $0.65<z\leq 0.80$. 
Although there are no visible differences in the $\delta$ distributions of the AGN and SF(control) samples, their similarity is confirmed only in one redshift bin ($\rm{0.65<z\leq0.80}$) by the KS test and skewness values.

	\begin{table}
		\centering                         
		\begin{tabular}{p{1.5cm} p{1.5cm} p{1.5cm}}    
			\hline 
			Redshift bin  & $\rm{S_{AGN}}$ & $\rm{S_{SF(control)}}$ \\
			\hline 
            \hline
			0.50$<$z$\leq$0.90 & 0.20$\pm$0.01 & 0.03$\pm$0.01   \\
            \hline
			0.50$<$z$\leq$0.65 & 0.25$\pm$0.01 & 0.06$\pm$0.01   \\
			0.65$<$z$\leq$0.80 & 0.08$\pm$0.01 & -0.04$\pm$0.02   \\
			0.80$<$z$\leq$0.90 & 0.42$\pm$0.13 & -0.31$\pm$0.11   \\
			\hline
		\end{tabular}
		\caption{Skewness of the $\delta$ distribution for the AGN and SF(control) samples in different redshift bins. 
			 }             
		\label{tables:skewness}     
	\end{table}

\subsection{Fraction-density relation}\label{sec:fraction_density_relation}

\begin{figure*}
	\centerline{\includegraphics[width=0.99\textwidth]{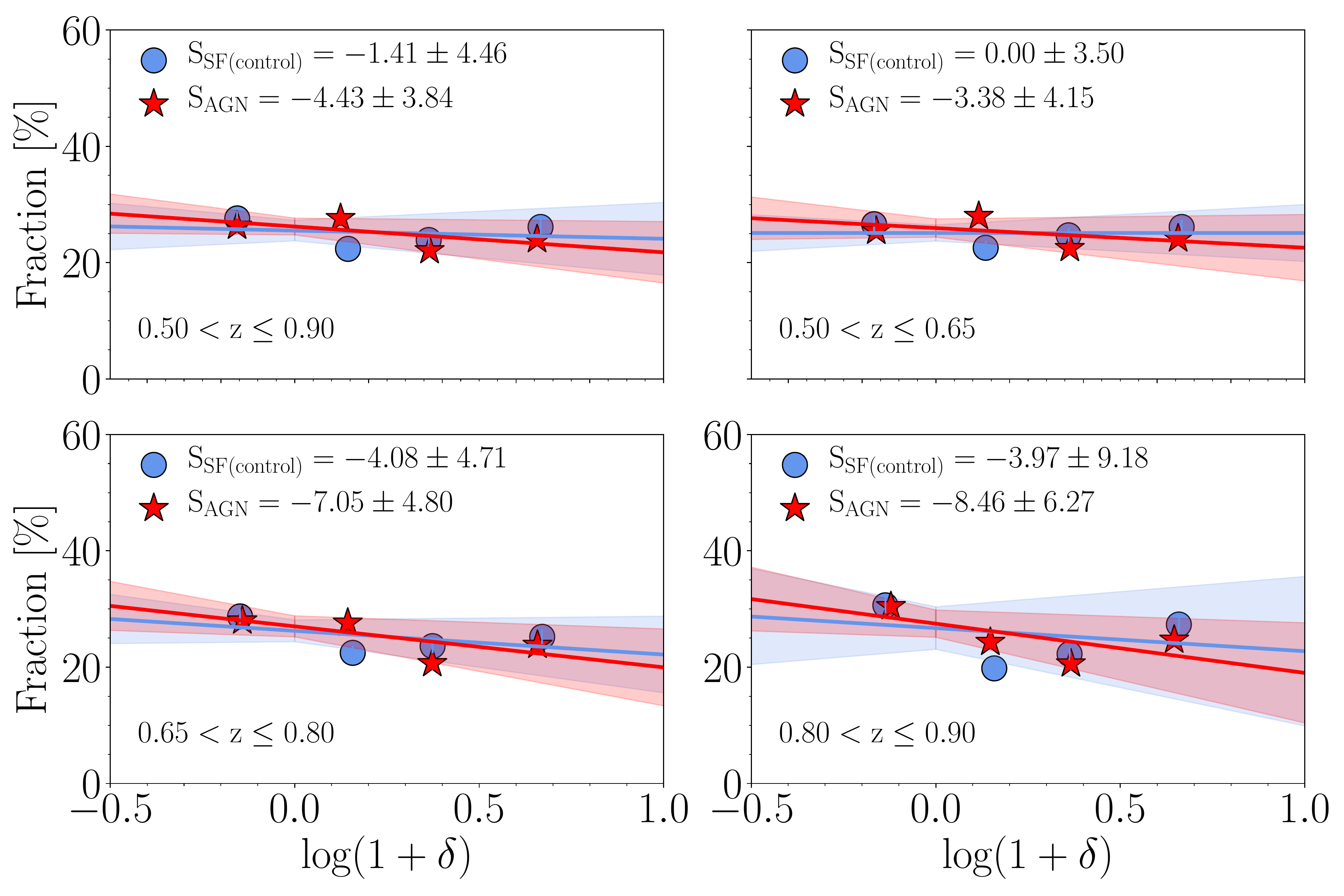}}
	\caption{In each panel the fractions of SF(control) and AGN galaxies as a function of the local density are given for three redshift bins and all together going clockwise. The $\delta$ bins correspond to the quartiles of the $\delta$ distribution of star-forming dwarf galaxies. The medians of $\delta$ in each bin for the SF(control) and AGN samples are marked with blue and red points, respectively. The solid lines correspond to the weighted fit. Shaded stripes around lines display $1\sigma$ of the fit. The slopes of the fit are given in the legend. }
	\label{fig:delta_fraction}
\end{figure*}

The relation between fraction of AGN and SF(control) galaxies and the environment is shown in Fig.~\ref{fig:delta_fraction}. 
As mentioned before, the fractions are weighted with selection weight $w$, and $1/Vmax$ method to account for sample/mass incompleteness. 
The $\delta$-bins correspond to the quartiles of the $\delta$ distribution of the 2,238 star-forming galaxies selected using emission line diagnostic diagram. 
Fractions are normalised to the total number of galaxies in each sample, i.e. the fraction sums up to 100\% for each sample. 
With such normalisation we are able to directly compare the slopes between different samples.  
In many works, when estimating $\delta$ percentiles and AGN fraction all galaxies are taken into account, i.e. including passive and star-forming objects~\citep[e.g.][]{Silverman2009, Hwang2012, Martini2013}. 
Defining $\delta$ percentiles and fraction relative to solely star-forming dwarf galaxies allows us to quantify the change purely in respect to hosting or not AGN, i.e. to investigate the intrinsic correlation between AGN and their environment~\citep{Man2019}. 

As it is clear from Fig.~\ref{fig:delta_fraction}, 
the SF(control) and AGN samples show mild negative fraction-$\delta$ relation within each redshift bin, with a slope three times larger for the AGN sample ($S=-4.43\pm3.84$) than for the SF(control) sample ($S=-1.41\pm4.46$), but consistent within $\sim1\sigma$. 
For both samples, slopes are at least doubled when moving to higher redshift, however, are constant within $<1\sigma$. 
The fraction of the SF(control) and AGN samples tends to be independent from the environment (within $<2\sigma$) and similar to each other (within $<1\sigma$) with no signs of significant evolution since z$=0.9$ (within $<1\sigma$).

\subsection{Dependence on redshift and stellar mass}\label{sec:z_fraction}
\begin{figure}
	\centerline{\includegraphics[width=0.49\textwidth]{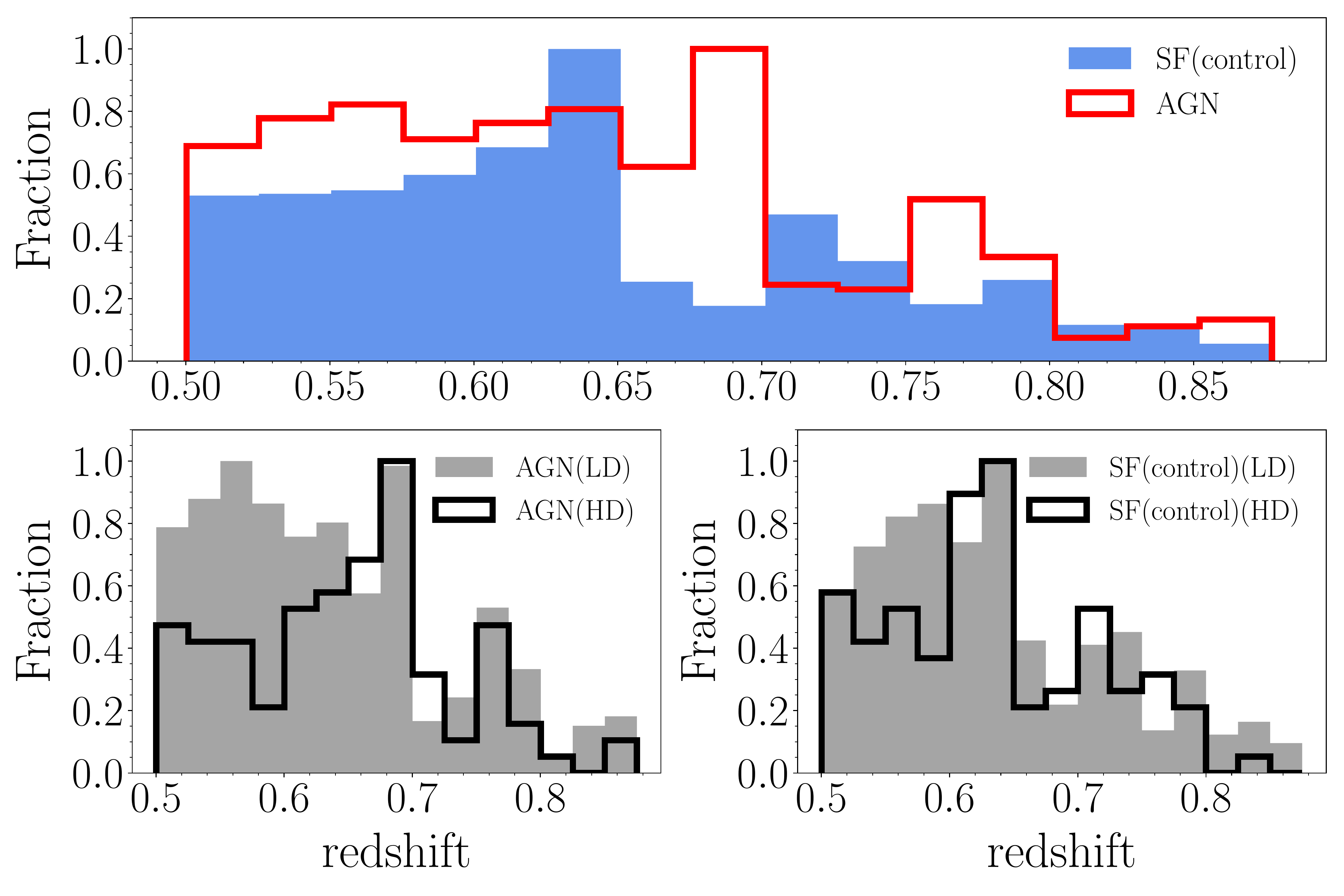}}
	\caption{Upper panel: normalised redshift distribution for the SF(control) and AGN samples. Lower panels: normalised redshift distributions in LD and HD environments for the AGN (left) and SF(control) (right) samples. 
	}
	\label{fig:z_fraction}
\end{figure}

\begin{figure}
	\centerline{\includegraphics[width=0.49\textwidth]{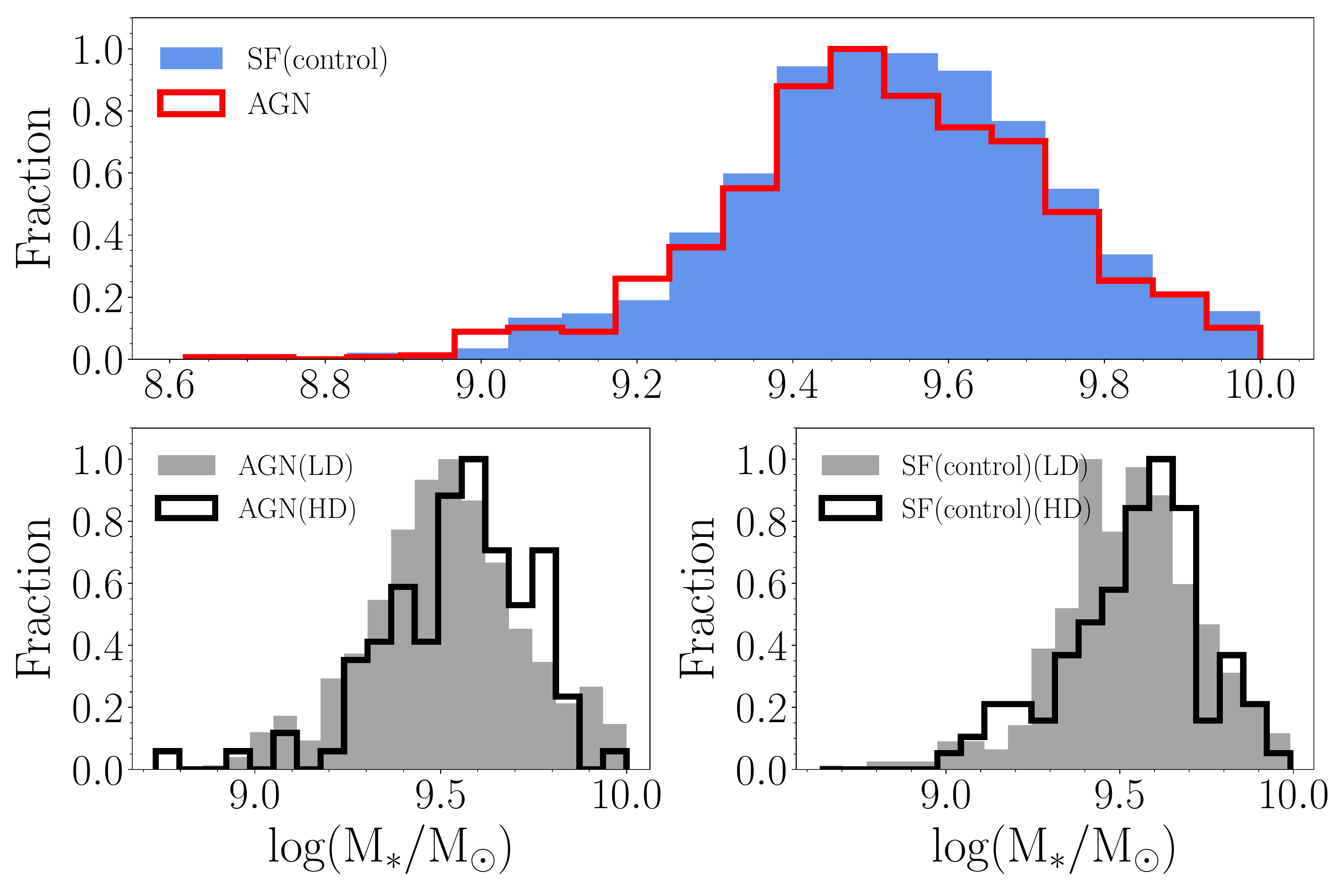}}
	\caption{Upper panel: normalised stellar mass distribution for the SF(control) and AGN samples. Lower panels: normalised stellar mass distributions in LD and HD environments for the AGN (left) and SF(control) (right) samples. 
	}
	\label{fig:mass_fraction}
\end{figure}

Although no strong evidence for the environmental dependence is found  for the AGN sample, there are some hints about some redshift discrepancies: smaller skewness of $\delta$ distribution and high p-value found in redshift bin $\rm{0.50<z\leq0.65}$, see Sec.~\ref{sec:density_distribution}); stronger negative trend of the fraction-$\delta$ relation at higher redshift (the slope is doubled but still in agreement within errorbars, see Sec.~\ref{sec:fraction_density_relation}).  
To investigate the possible evolution with redshift, we check if the redshift of the AGN and SF(control) distributions are uniformly distributed. 
As clearly shown in upper panel in Fig.~\ref{fig:z_fraction} the fraction of AGN and SF(control) galaxies is higher at the lowest redshift bin, $\rm{0.50<z\leq0.65}$, and lower at the highest redshift bin, $\rm{0.80<z\leq0.90}$. 
Interestingly, the intermediate redshift bin, $\rm{0.65<z\leq0.80}$, is characterised by strong peaks: at $\rm{z\sim0.68}$ seen for the AGN sample and at $\rm{z\sim0.64}$ for SF(control) sample (see upper plot in Fig.~\ref{fig:z_fraction}). 
As shown in the lower left panel in Fig.~\ref{fig:z_fraction}, the AGN fraction-redshift relation shows mild dependence on the environment, as low-redshift AGN reside in LD environments, while higher-redshift AGN are found both in LD and HD environments. This might be explained by the fact that galaxies in clusters and pairs at higher redshift are not anymore in structures at lower redshift (i.e. higher-redshift galaxies residing in HD environment move to a LD environment at lower redshift).   
Similarly, SF(control) galaxies residing in LD environments are observed rather at lower redshift ($\rm{0.50<z<0.65}$), whereas SF(control) galaxies residing in HD environments are distributed more uniformly with a peak at $\rm{z\sim0.60-0.65}$. 
This suggests that the SF(control) and AGN samples show similar redshift distributions, however, the peak is shifted to slightly lower redshifts for the SF(control) sample. 
At the same time, the stellar mass distributions for the SF(control) and AGN samples are similar as shown in the upper panel in Fig.~\ref{fig:mass_fraction} and is nearly independent on the environment (with a 0.02 dex higher mass for galaxies found in HD environments; see lower panels). 
The higher fraction of dwarf galaxies since $\rm{z\sim0.7}$ might be a consequence of turning on the AGN activity at this epoch or the consequence of the survey incompleteness, as fainter galaxies at higher redshift dropped out from the sample (with AGN observed at slightly higher redshift than non-AGN dwarfs as the AGN light contribution makes them brighter). 
However, such possible survey incompleteness bias should be addressed by the selection of three complete K-band luminosity redshift bins (see Sec.~\ref{sec:sample_selection}) and the introduction of the survey weight $w$ and $1/V_{max}$ method (see Sec.~\ref{sec:results}). 
The increase in dwarf galaxies since $\rm{z\sim0.6-0.7}$ was also reported by~\cite{moutard2018}, who found that the reservoir of low-mass star-forming galaxies located in very dense regions has grown between $\rm{z\sim0.6}$ and z $\sim0.4$. 
The increase AGN fraction since $z\sim0.7$ could be connected to the AGN feedback in suppressing star formation activity~\citep[e.g.][]{Manzano-King2019,Davis2022,Koudmani2022}. 

\subsection{NUVrK diagram}\label{sec:nuvrk}

In this Section, we analyse the distribution of the SF(control) and AGN samples in the NUVr-rK ($\rm{NUVrK}$) diagram~\citep{arnouts}. 
The $\rm{NUVrK}$ diagram is a colour-colour space commonly used to separate different galaxy types: passive (red), intermediate (green), and active (blue)~\citep[e.g.][]{arnouts,fritz14,moutard16b,davidzon16,siudek17,siudek18,Siudek2022}. 
Thanks to the sensitivity of NUVr colour to the recent star formation and rK to stellar ageing and dust attenuation, the NUVrK diagram is very well suited to distinguish SFHs characterised by different quenching time-scales. 
Namely, old, quiescent galaxies exhibit redder $\rm{NUVr}$ colours, while galaxies characterised by younger stellar ages are bluer. 
However, as $\rm{NUVr}$ colour is sensitive to dust attenuation, dusty star-forming galaxies may also show reddened $\rm{NUVr}$ colours. 
This degeneracy in the $\rm{NUVr}$ colour can be broken with the $\rm{rK}$ colour, which is less sensitive to dust obscuration~\citep{arnouts07,martin07}. 

\begin{figure*}
	\centerline{\includegraphics[width=0.99\textwidth]{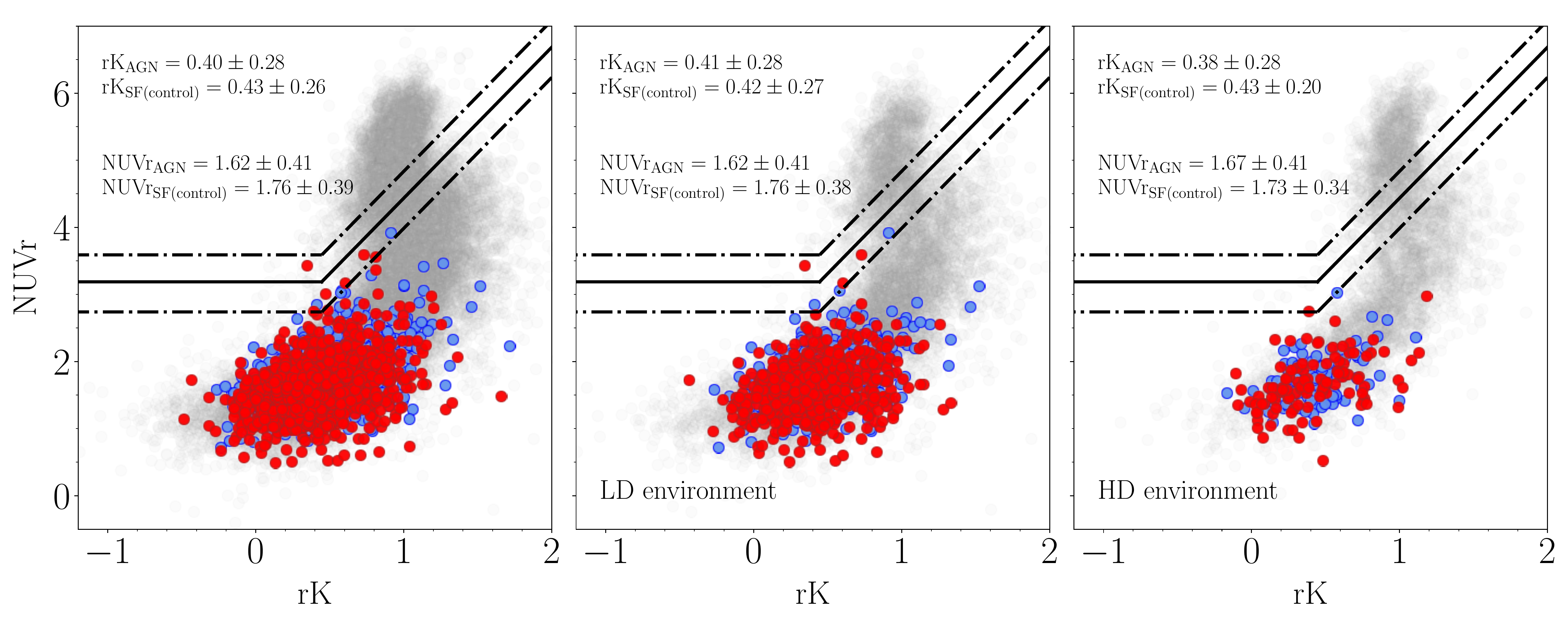}}
	\caption{NUVrK for VIPERS galaxies from \citealt{Siudek2022} (in gray), SF(control, blue diamonds) and AGN (red circles) samples. The mean and standard deviation of the rK and NUVr colours for the SF(control) and AGN galaxies are shown in the upper left corner. The solid line corresponds to the separation of red passive and blue star-forming galaxies. The dashed lines enclose the area of the green valley galaxies as proposed by~\citealt{moutard16b}.}
	\label{fig:nuvrk}
\end{figure*}

Figure~\ref{fig:nuvrk} shows the NUVrK diagram for the SF(control) and AGN samples (left panel) residing in LD (middle panel) and HD (right panel) environments on the background of 31,631 VIPERS galaxies (including also massive and red galaxies) used to study the environments of different types of galaxies by \cite{Siudek2022}. 
AGN and SF(control) galaxies are distributed similarly on the NUVrK diagram, occupying the region of star-forming galaxies according to the classification introduced by~\cite{moutard16b} with a negligible number of galaxies in LD environment found in the green valley.  
SF(control) galaxies tend to have tails towards redder rK colour, which are characterising edge-on galaxies~\citep{arnouts, moutard16b}. 
This displacement is, however, not statistically significant, as the mean NUVr and rK colours are in agreement within $0.5\sigma$. 
For all samples their location on the NUVrK diagram tend to be independent on the environment.

\subsection{4000 \AA~break strength}\label{sec:d4000}

\begin{figure}
	\centerline{\includegraphics[width=0.49\textwidth]{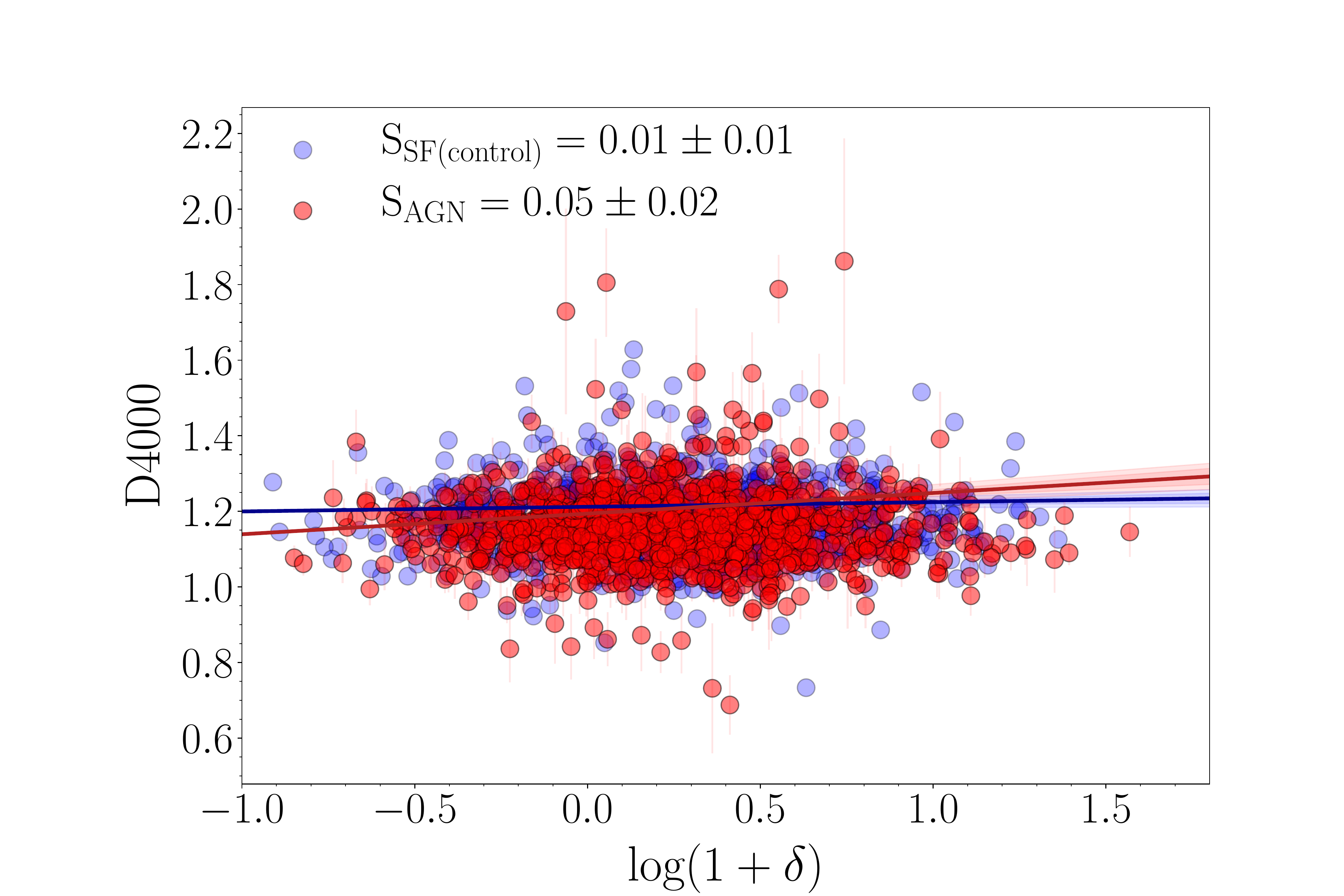}}
	\caption{The D4000-$\delta$ relation for the SF(control) and AGN samples. The solid line corresponds to the weighted fit. Shaded stripes around lines display $1\sigma$ of the fit. The slope of the fit is given in the legend.}
	\label{fig:D4000}
\end{figure}

The 4000 \AA~break strength (D4000) is one of the most important features in the optical part of a galaxy spectrum.  
We adopt the narrow definition of the D4000 proposed by~\cite{Balogh1997} as the ratio of the flux before (the red continuum; 4000 - 4100~\AA) to that after the break itself (the blue continuum; 3850 - 3950~\AA). 
In this narrow wavelength range the absorption of several ionised metallic elements are accumulated. 
In young, hot stars these elements are multiply ionised and the line opacities decrease, which is reflected in a small D4000, while the D4000 becomes larger for older stellar populations. 
Thus, D4000 is widely used in galaxy evolution studies as a stellar age indicator~\citep[e.g.][]{balogh1999,kauffmann03, siudek17}. 
The D4000-$\delta$ relation for the SF(control) and AGN samples is shown in Fig.~\ref{fig:D4000}. 
For both samples the relation is constant within $2\sigma$, suggesting that D4000 is independent on the environment for both samples. 
This indicates that not only dwarf galaxies hosting AGN are characterised by similar stellar ages as the ones non-hosting AGN, but also that the stellar populations of dwarfs residing in LD environments are of the same age as the ones found in HD regions.   
Moreover, on average, the D4000 found for both samples (D4000~$\sim1.2$) is typically found for young blue galaxies ($\rm{D4000<1.5}$ as defined by~\citealt{kauffmann03}) further supporting their classification as blue, star-forming galaxies (see Sec.~\ref{sec:nuvrk}). 

\subsection{Galaxy size}\label{sec:re}
\begin{figure}
	\centerline{\includegraphics[width=0.49\textwidth]{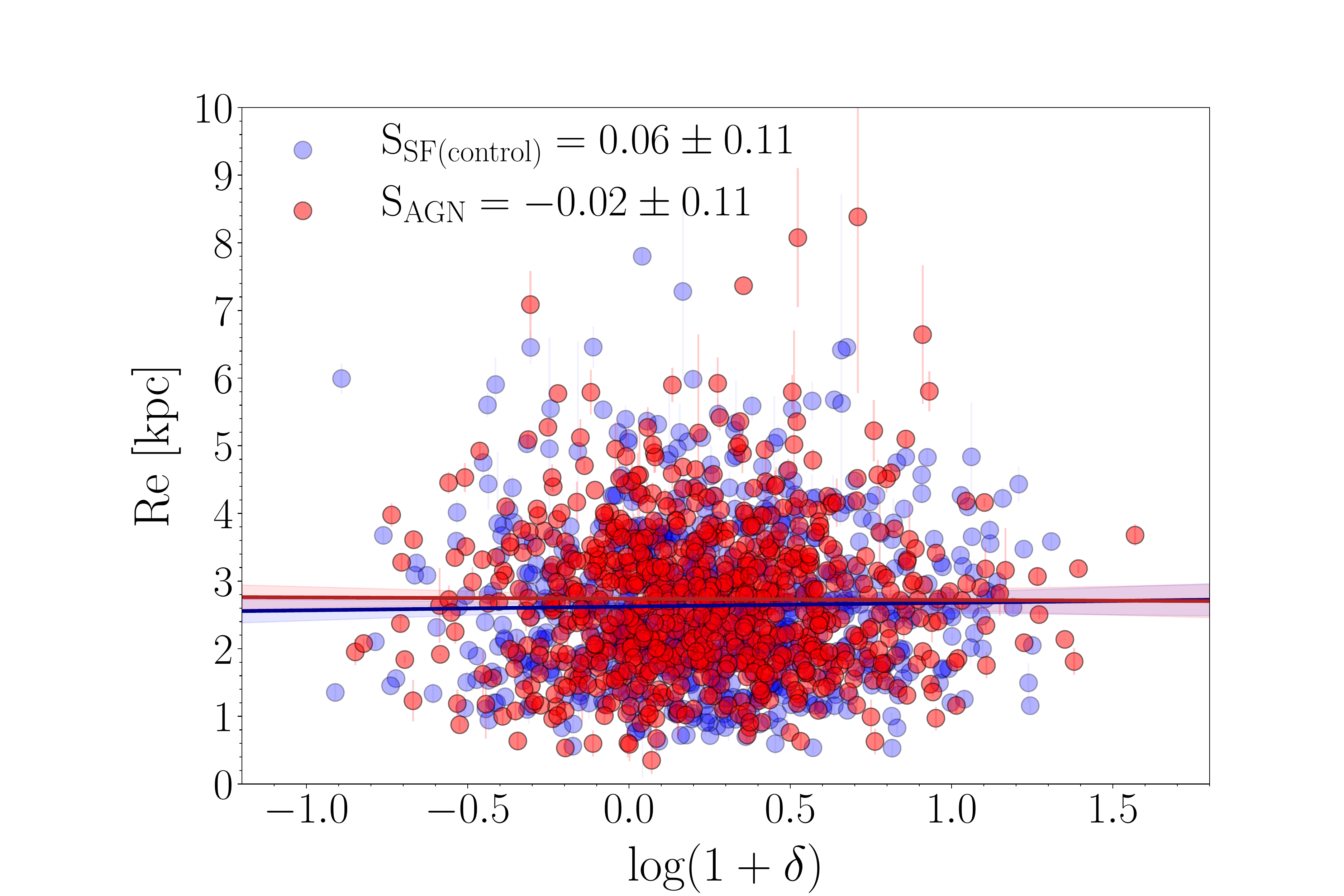}}
	\caption{Re-$\delta$ relation for the SF(control) and AGN samples. The solid line corresponds to the weighted fit. Shaded stripes around lines display $1\sigma$ of the fit. The slope of the fit is given in the legend.}
	\label{fig:Re}
\end{figure}

The morphology of galaxies was found to be correlated with the environment~\citep[e.g.][]{Dressler1980, Siudek2022}. 
In particular, the structure of dwarf galaxies should reflect the environments in which they reside. 
Dwarfs in HD environments are affected by ram pressure stripping~\citep[e.g.][]{Gunn1972,Grebel2003} or tidal stirring~\citep[e.g.][]{Mayer2001,Kazantzidis2011} on top of the cluster effects~\citep[e.g. galaxy harassment,][]{Moore1998,Smith2015}.  
Dwarf galaxies in LD environments, untouched by these effects, are shaped rather by stellar feedback~\citep{Geha2012}.  
\cite{Carlsten2021} found that dwarf galaxies in cluster environments tend to be slightly larger than dwarf galaxies in the field at a fixed stellar mass. 
Those authors associated the observed increase of size (10\%) to more intense tidal stripping and heating of galaxies in extreme cluster environments, which aligns with theoretical expectations. 
It can be seen from Fig.~\ref{fig:Re} that there is no significant ($<1\sigma$) difference in the $R_{\mathrm{e}}$-$\delta$ relation between galaxies hosting or not AGN. 
Moreover, the trend of the $R_{\mathrm{e}}$-$\delta$ relation is constant within $0.5\sigma$ for both samples. 
This suggests that hosting AGN does not affect the size of dwarf galaxies and that the role of environment is also negligible in shaping dwarf galaxies. 

\subsection{Main sequence}\label{sec:MS}

\begin{figure*}
	\centerline{\includegraphics[width=0.99\textwidth]{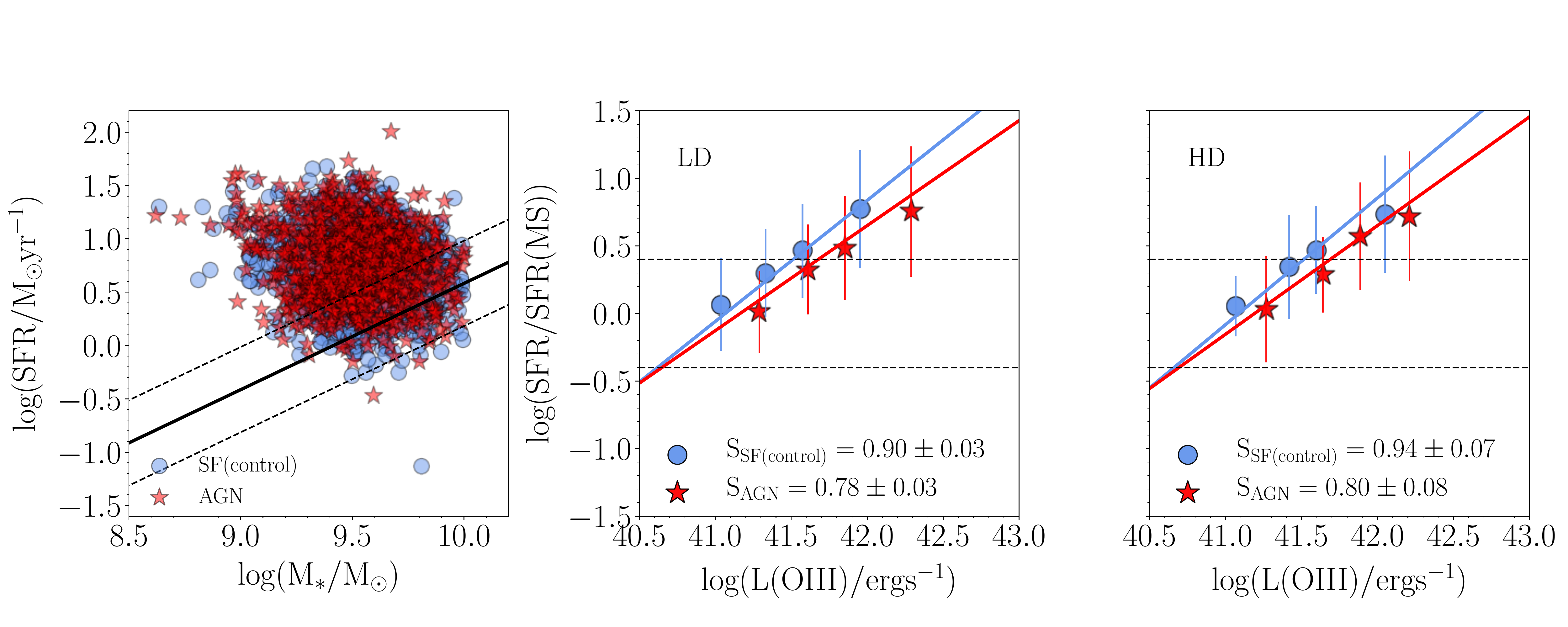}}
	\caption{Left panel: SFR(OII)-$\rm{M_{*}}$ relation for the SF(control) and AGN samples. The black solid line corresponds to the MS trend at z$=0.64$ as found by~\citealt{Schreiber2015}. Middle panel: The SFR distance-[OIII] luminosity found for the SF(control) and AGN samples in LD environments. The lines correspond to the linear fits of the relation found based on the SFR(OII). The points indicate the means of SFR(SED) found in luminosity bins. The loci of starburst and passive are delimited by the dashed lines ($\pm$0.4 dex). Right panel: the same as the middle one but for HD environments.}
	\label{fig:MS}
\end{figure*}

The SFR-$\rm{M_{*}}$ relation, called the galaxy main sequence (MS), is driven by physical mechanisms regulating galaxy growth and gas accretion~\citep[e.g.][]{Bouche2010}. 
As shown in the left panel of Fig.~\ref{fig:MS}, the bulk of SF(control) and AGN galaxies occupy the MS and starburst region. 
The MS relation is derived at the mean redshift of our sample, z $=0.64$, from~\cite{Schreiber2015} after rescaling by a factor of 1.7 to account for the usage of different IMF during the SED fitting (~\citealp{Salpeter1955} vs. ~\citealp{chabrier}). 
Similarly,~\citealt{Vietri2022} found that VIMOS AGN Type II are placed in the MS and starburst locus.
However, they do not report such a high fraction of AGN above the MS (see Fig. 4 in~\citealt{Vietri2022}) as that observed in our sample. 
This might be a consequence of the sample selection, as our SFR(OII) was derived from the [OII]$\lambda3726$ line corrected for dust extinction (see Sec.~\ref{sec:SFR(OII)}), whereas~\citealt{Vietri2022} used uncorrected [OII]$\lambda3726$ fluxes. 
Moreover,~\citealt{Vietri2022} removed the AGN contribution from SFR(OII) by subtracting 10\% of the L(OIII)~\citep[][]{Zhuang2019}, assuming that the high ionisation lines are powered by AGN activity~\citep{Kauffmann2003b}.
As we are interested in how AGN activity can impact the SFR, we do not apply this correction in our sample, what may explain the higher SFR found for our sample than the one presented in~\citealt{Vietri2022}. 

As it is clear from Fig.~\ref{fig:MS}, we do not see any difference in the MS relation between the AGN and SF(control) samples, suggesting that the AGN feedback on the star formation activity in dwarf galaxies is negligible. 
To further explore the impact of AGN on the SFR and its potential role in quenching star formation with increasing AGN luminosity~\citep[e.g.][]{Hopkins2016}, we investigate the distance from the MS as a function of AGN power (given as L(OIII), see Sec.~\ref{sec:SFR(OII)}). 
The distance to the MS from~\cite{Schreiber2015} as a function of L(OIII) is shown in the middle and right panels in Fig.~\ref{fig:MS} for LD and HD environments, respectively. 
To account for a possible bias introduces by the choice of the SFR estimations, we plot SFR(SED; marked with points) and SFR(OII; marked with lines). 
As clearly shown in the Figure, the relations are independent on the methodology used to derive SFRs. 
Our findings are in agreement with~\cite{Vietri2022}, who found that the majority of galaxies (88\%) located in and above the MS are dwarf galaxies hosting narrow-line AGN and that the distance to the MS increases with increasing L(OIII). 
The trend of the relation shows a mild negative offset for the AGN sample and is the same within $2(1)\sigma$ for LD(HD) environments. 
This offset could point to a role of AGN in suppressing star formation activity in dwarf galaxies, however it is not statistically significant.

\subsection{Post-starburst phase}\label{sec:PSB}
\begin{figure*}
	\centerline{\includegraphics[width=0.99\textwidth]{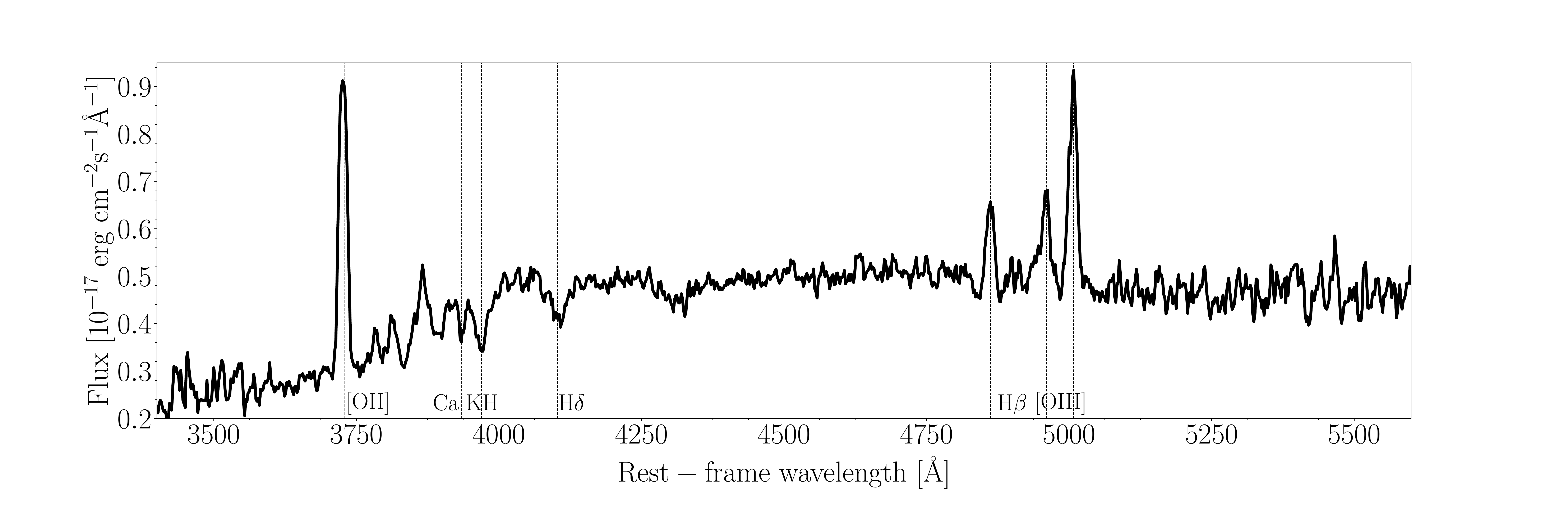}}
	\caption{Stacked spectrum of the 25 post-starburst AGN dwarf galaxies. The [OII], CaII H and K lines, H$\delta$, H$\beta$ and [OIII] are marked with black lines. }
	\label{fig:PSBspectrum}
\end{figure*}

Recent studies have linked AGN with post-starburst (E+A) galaxies (e.g. \citealt{Melnick2015}; \citealt{Baron2017}). AGN with post-starburst hosts may represent an intermediate phase, i.e galaxies in transit between the starburst stage and the fully quenched post-starburst stage, and may confirm whether AGN and starburst activity co-evolve (\citealt{Wei2018}). We search for post-starburst AGN dwarf galaxies assuming that they are simultaneously showing the signature of AGN (i.e. emission lines and/or X-ray/radio emission, see Sec.~\ref{sec:data}) and post-starburst stellar properties. There are different methods to select post-starburst galaxies, all of which intend to select galaxies that experienced rapid (< 100 - 200 Myr) decline in their recent ($\rm{\lesssim 1 Gyr}$) starburst (e.g \citealt{French2021}). 
In order to determine the post-starburst signature we used the absolute magnitudes in U, B, V filters. Following the approach proposed by \cite{Suess2022}, we identify 25 post-starburst AGN candidates (2\% of AGN sample) 
with $\rm{U-B}>0.975$ and $-0.25< B-V<0.45$. 
Due to VIPERS resolution, the H$\delta$ line (standardly used to identify post-starburst phase) is not distinguishable in individual VIPERS spectra, but can be detected if at least 20 spectra are stacked (\citealt{siudek17}). We thus stack the 25 post-starburst galaxy spectra by normalising the rest-frame spectra with the median flux computed at 3700-4200 \AA, and applying a median stacking (see more details about stacking procedure in \citealt{siudek17}). 
The stacked spectrum shown in Fig.~\ref{fig:PSBspectrum} reveals indeed signs confirming their post-starburst nature: a relatively flat continuum, Balmer high-order absorption lines, an inverted CaII H/K line ratio, and a strong H$\delta$ line ($\rm{EW(H\delta)}=6.03$~\AA). 
As expected, the 25 post-starburst AGN dwarf galaxies have lower SFR ($\rm{SFR(SED)=0.33\pm0.39}$) than the entire AGN sample and a majority (68\%) of them are placed on or below the MS.  
At the same time, they are more preferably found in LD environments (40\%) than in HD regions (15\%). Similarly, \cite{Yesuf2022} found that only $\sim2-4\%$ of AGN with $\rm{H\delta>4}$~\AA~live in clusters.  
As selection of a pure and complete sample of post-starburst galaxies is extremely challenging (e.g. \citealt{Baron2022}), in a future work we plan to study the relationship between post-starburst AGN and normal AGN based on a sample where at least $H\delta$ is detected on individual spectra. 

\subsection{The black hole mass}~\label{sec:BHmass}

The black hole mass for the galaxies in the AGN sample is derived using Eq. 1 in~\cite{Ferre-Mateu2021MNRAS.506.4702F}, which was obtained by combining Eqs. 1, 5 and 6 of \cite{Baron2019MNRAS.487.3404B}:

\begin{equation}
\label{eq1}
\rm
log M_{BH} = log \epsilon + 3.55log(L_{[OIII]}/L_{H\beta}) + 0.59log L_{bol}-20.96 
\end{equation}

where $L_\mathrm{[OIII]}$ and $L_\mathrm{H\beta}$ are the extinction-corrected luminosities of the narrow [OIII]$\lambda5007$ and H${\beta}$ emission lines and $L_\mathrm{bol}$ is the bolometric luminosity derived as in \cite{Netzer2009MNRAS.399.1907N}:

\begin{equation}
\rm
log L_{bol} = log L_{H\beta} +3.48 + max[0,0.31(log [OIII]/H\beta -0.6)]
\end{equation}

We assume a scale factor $\epsilon$=1.075 as in \cite{Baron2019MNRAS.487.3404B}. 
We find a range of black hole masses peaking at log $\rm{M_{BH}}$ = 6 - 10 M$_{\odot}$ and with a median of log $\rm{M_{BH}}$ = 8.2 M$_{\odot}$, in agreement with the distribution found by \cite{Vietri2022}. Dwarf galaxies with log $M_\mathrm{BH} >$  6 M$_{\odot}$ are unusual (e.g. \citealt{Reines2015ApJ...813...82R}; \citealt{Greene2020}) and would be overmassive with respect to the black hole-galaxy scaling relations, even if there is a flattening in the low-mass regime (e.g.~\citealt{Mezcua2017}, \citealt{Martin-Navarro2018}). Such overmassive black holes have however been recently found at $\rm{z\gtrsim 0.4}$ by Mezcua et al. (submitted), in agreement with cosmological dwarf zoom-in simulations that predict the existence of a population of overmassive black holes in dwarf galaxies powered by efficient AGN activity~\citep{Koudmani2022}.  

\begin{figure}
	\centerline{\includegraphics[width=0.49\textwidth]{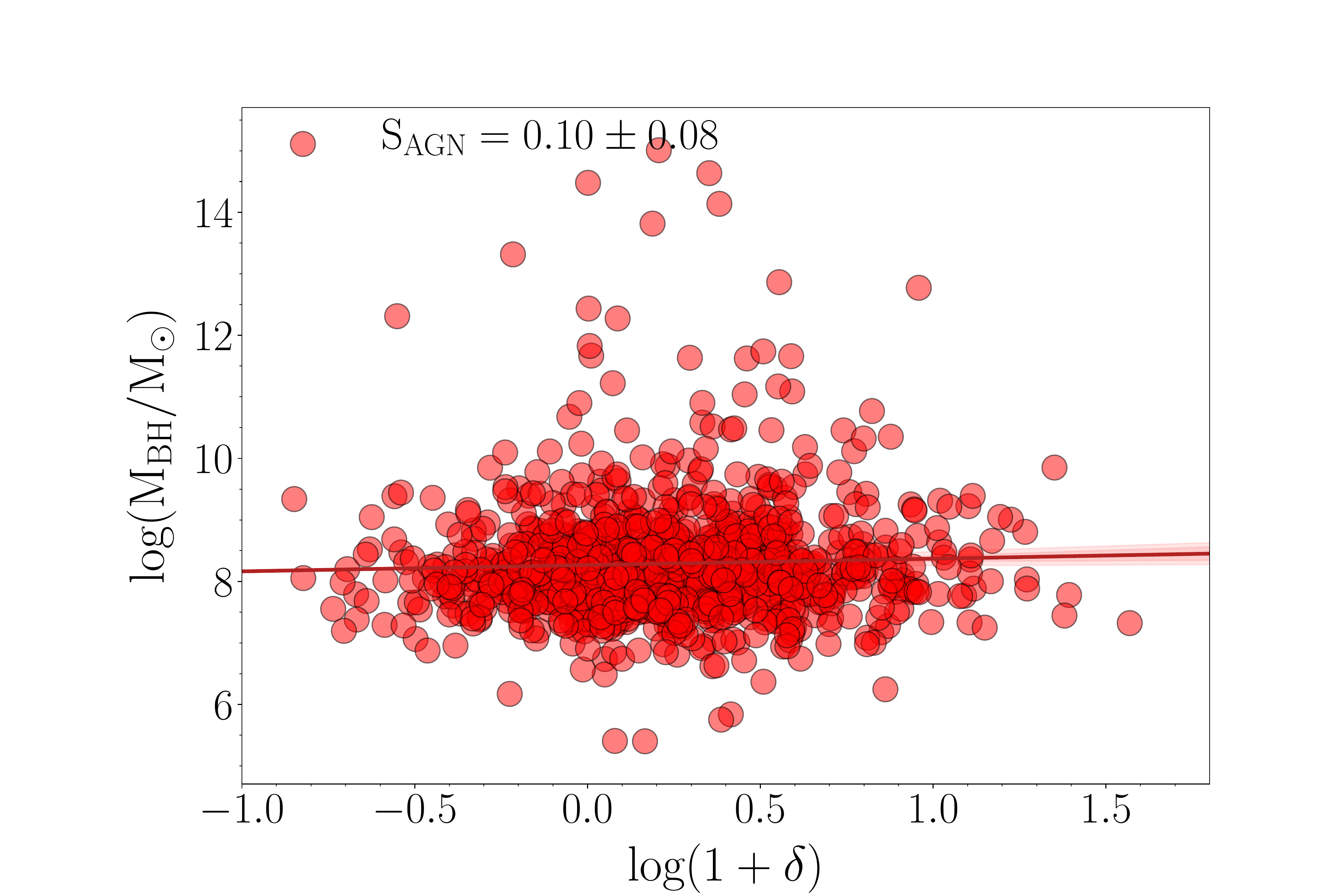}}
	\caption{Black hole mass-$\delta$ relation for the AGN sample. The solid line corresponds to the linear fit. The shaded area indicates the $1\sigma$ uncertainty of the fit. }
	\label{fig:BHmass}
\end{figure}

The $\rm{M_{BH}}$-$\delta$ relation is shown in Fig.~\ref{fig:BHmass}. 
As it is clear from this plot, BH masses are independent on the environment. 

\section{Discussion}\label{sec:discussion}

VIPERS dwarf galaxies are highly star-forming galaxies (preferentially located in the MS and starburst region, see Fig.~\ref{fig:MS}) in agreement with previous studies of dwarf galaxies hosting AGN~\citep[e.g.][]{Mezcua2016,Mezcua2018a,Mezcua2020, Vietri2022}. 
They are preferably found in LD environments, as expected for dwarf star-forming galaxies~\citep{davidzon16, Cucciati2017, Siudek2022}. 
We do not find any significant difference in the $\delta$ distribution of AGN and SF(control) galaxies (see Tab.~\ref{table:number_in_HD_LD} and Figs.~\ref{fig:delta_distribution} and~\ref{fig:delta_fraction}) indicating that they reside in similar environments.  
AGN and non-AGN dwarf galaxies show also similar physical, properties which do not evolve when moving to denser environments (see Figs.~\ref{fig:mass_fraction},~\ref{fig:nuvrk},~\ref{fig:D4000},~\ref{fig:Re}, and~\ref{fig:MS}). 
For the AGN sample the BH mass is also independent on the environment (see Fig.~\ref{fig:BHmass}). 

The overabundance of dwarf galaxies in LD environment is in line with the theoretical expectations, as the most important factor in fuelling AGN activity is having a supply of gas to feed the core~\citep{Sabater2013}. 
The cold gas content in dwarf galaxies is more vulnerable than in more massive galaxies due to their shallow gravitational potential wells. 
The cold gas reservoirs in dense local environments is stripped and heated, so we may expect LD environments to be dominated by low-mass galaxies. 

Our findings of dwarf galaxies residing in similar environments whether or not hosting AGN are in agreement with studies presented by~\cite{Kristensen2020}. 
Using 62,258 dwarf galaxies with ${\rm log(M_\mathrm{*}/M}_{\odot})\lesssim10.70$ they do not found any environmental difference for dwarf galaxies hosting or not AGN in the local Universe. 
However, the authors stress that no differences in environment between AGN and non-AGN host dwarf galaxies may be  falsified by biases in AGN selection, mass trends or other factors. 
In a follow-up work using Illustris simulations,~\cite{Kristensen2021} find a preference for HD environments for dwarf galaxies non-hosting AGN, indicating a non-negligible role of mergers and environment in triggering AGN activity.    
However, the authors report that the overabundance of non-AGN in HD environments is at least partially driven by red star-forming galaxies, which are a population not observed by recent surveys or missed by the selection procedures (see Fig.~\ref{fig:nuvrk} where no dwarf galaxies are located in the red star-forming region, i.e. with $\rm{rK\gtrsim1.5}$). 
From an observational point of view,~\cite{Manzano-King2020} further confirm that dwarf galaxies hosting or non-hosting AGN reside in similar environments using 45 AGN and 19 non-AGN dwarf galaxies observed with Keck LRIS longslit spectroscopy. 
Despite the negligible role of the environment, the authors suggest that AGN influence the gas kinematics and suppress star formation in dwarf galaxies. 
A recent study of 78 radio AGN in dwarf galaxies observed at $\rm{0.1<z<0.5}$ also finds no discernible differences between the environments of AGN and non-AGN dwarf galaxies~\citep{Davis2022}.   
Moreover, those authors do not find differences in the SFR or recent interactions (based on the incidents of tidal features) for AGN and non-AGN samples, suggesting that no special circumstances are needed for the presence of the AGN. 

\begin{figure}
	\centerline{\includegraphics[width=0.49\textwidth]{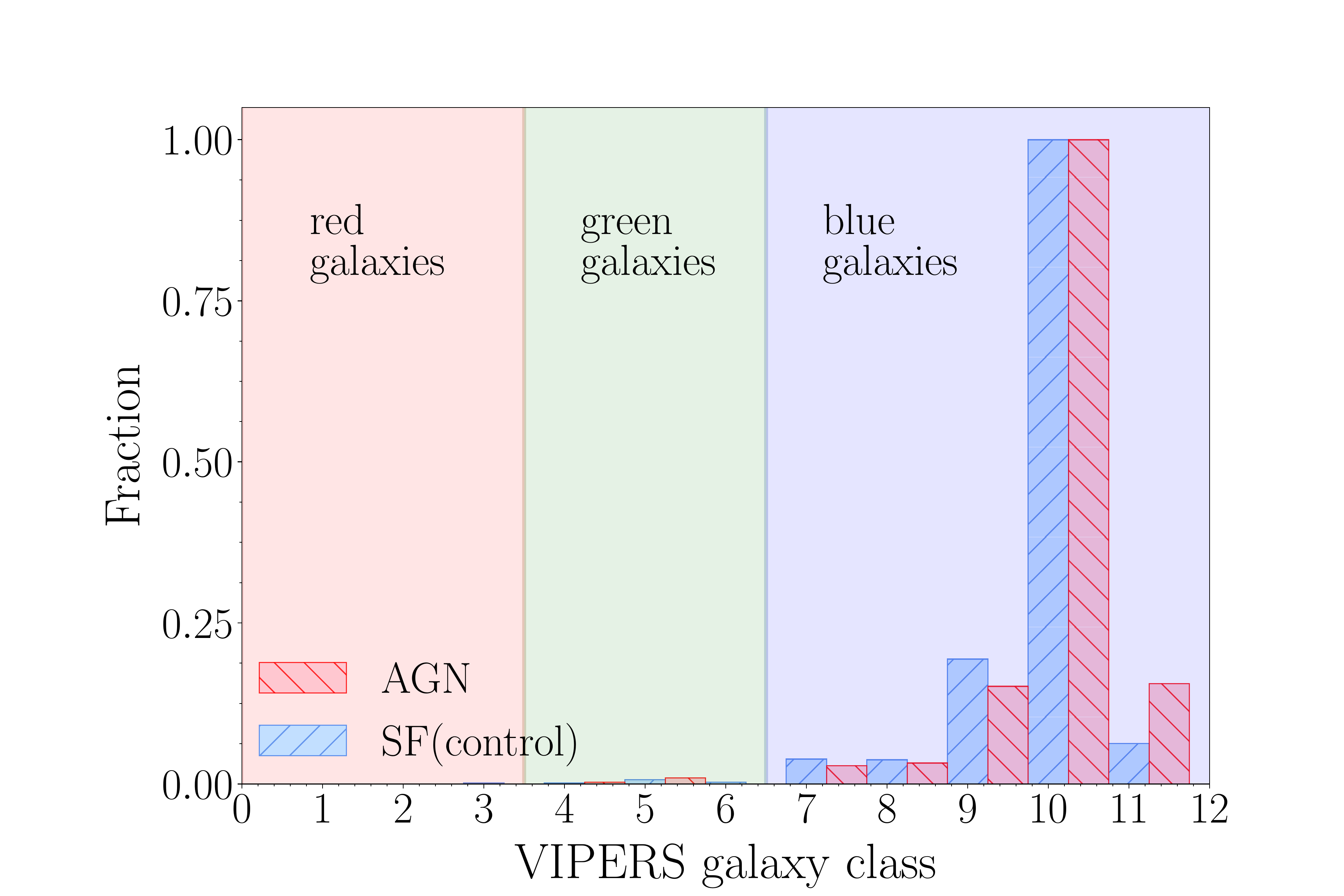}}
	\caption{Distribution of galaxy classes (as described in~\citealt{siudek18}) for the AGN and SF(control) dwarf galaxies normalised to the total number of AGN and SF(control) galaxies, respectively.}
	\label{fig:FEMclass}
\end{figure}

The comparison of our AGN and SF(control) samples shows that the D4000, $R_{\mathrm{e}}$, and colours (NUV, rK and ri) are consistent with each other. 
To further confirm the similarity of the AGN and SF(control) galaxy properties, we use a machine-learning based classification in multi-dimensional space introduced by~\citealt{Siudek2018a,siudek18} (see also~\citealt{Siudek2022} and Siudek et al. submitted).  
In Fig.~\ref{fig:FEMclass} we show the distribution of the AGN and SF(control) samples among the different galaxy classes proposed by~\cite{siudek18}. 
The number of AGN(SF(control)) galaxies in each class is normalised to the number of galaxies of the AGN(SF(control)) sample. 
The bulk of AGN and SF(control) galaxies is gathered in a class characterised by star-forming galaxies with the bluest NUVr and rK colours~\citep{siudek18}. 
More importantly, this class is composed mostly by dwarf galaxies (with log$\rm{(M_\mathrm{*}/M_{\odot})=9.56}$, i.e. in a dwarf-galaxy regime, \citealt{siudek18}). 
Automatically-selected class of dwarf galaxies is an example of advantages of adapting machine-learning tools for galaxy classification~\citep[e.g.][Siudek et al. submitted]{Siudek2018a,siudek18,turner2021,Siudek2022, Lisiecki2022}. At the same time, the VIPERS classification providing the completeness (27\%) and accuracy (at least 67\%) of selecting dwarf AGN is challenging the commonly used mid-infrared colours AGN selection (see details in Siudek et al. submitted) . 
At the same time, only a small fraction of the AGN and SF(control) samples is assigned to classes gathering green galaxies, as also confirmed by their position on the NUVrK diagram (see Sec.~\ref{sec:nuvrk}). 
This implies that our dwarf galaxy sample consists mainly of blue star-forming galaxies~\citep[][]{siudek18}. 
The consistency of the SFR and colours (as a proxy to gas richness) between the AGN and control samples suggests that AGN are not gas-enriched due to environmental effects (supply of gas) and do not show evidence for suppressing star formation. 
However, the SF(control) sample shows stronger offset from the MS than the AGN sample, which might suggest that AGN suppress star formation at some non-negligible level. This result is, however, not statistically significant ($\sigma<0.5$).

Summarising, our analysis further strengthens the finding that the triggering of AGN activity is independent on the environment or gas fraction (given as the SFR and colours) or sizes (as given by $R_{\mathrm{e}}$) and is universal since $\rm{z=0.9}$. 
Nevertheless, the negligible role of the environment might by biased by the unrepresentative sample of dwarf galaxies, as recent surveys may catch only the bright population of AGN with high star formation. 
Also, mechanisms triggering AGN activity might not occur at the same time when AGN activity is observed~\citep[e.g][]{Hopkins2012,Pimbblet2013,Kristensen2020}. Another challenge is the selection of pure and complete AGN and star-forming samples, especially in the dwarf galaxy mass regime. Methods based on the emission in different wavelengths (emission lines, infrared colours, X-ray and radio emissions) do not lead to the same samples and suffer from the contamination (e.g. \citealt{Hviding2022}). Tracing the redshift evolution is even harder as the most standard AGN selection method based on the optical BPT diagram is restricted to low-redshift ($\rm{z<0.45}$) samples (see Sec.~\ref{sec:BPT}). To extend the AGN study to intermediate-redshift ($\rm{z\sim1}$) we can take advantage of the emission line diagnostic diagram proposed by \cite{2010Lamareille} and used in this work; however, the BPT and \cite{2010Lamareille} diagrams result in slightly different samples. 
The comparison of the \cite{2010Lamareille} emission line diagnostic diagram with the BPT diagram (\citealt{Kewley2006}) was raised in \cite{2010Lamareille}. The accuracy is 99.7\% for star-forming galaxies, 59\% for Seyfert 2 (raising to 86\% if we include composite region) and 99.1\% for LINERS.  
The contamination is 16\% for star-forming galaxies and is negligible for Seyfert 2 galaxies, while LINERS are contaminated by composite objects. The BPT-selected composite objects (that reside between \citealt{Kewley2001} and \citealt{Kauffmann2003b} lines) are classified as star-forming galaxies (85\%) or LINERS (16\%) in the emission line diagnostic diagram proposed by \cite{2010Lamareille}. 
We consider the limitations of the purity of LINERS by showing that excluding LINERS from the VIPERS AGN sample does not affect our results (see App.~\ref{app:sanity_check}). To address the non-negligible contamination (by 16\%) of the star-forming sample, we compare the AGN and SF(control) samples with the classification based on the MEx diagram (see Sec.~\ref{sec:BPT} and App.~\ref{app:sanity_check}). We find that 90\% of SF(control) galaxies are star-forming on the MEx diagram, and incorporating the MEx-based selection does not affect our findings. 
Although we show that selection limitations do not affect our results (see App.~\ref{app:sanity_check}), it is clear that an efficient tool to identify AGN in dwarf galaxies is highly needed. 

Due to these observational limits, the low-mass regime is still rather unexplored, especially the role of AGN feedback on the star formation of the host. Our findings invite further investigation of low-mass galaxies hosting AGN and show the potential for future surveys (e.g. DESI, LSST) to even expand the population of dwarf galaxies hosting AGN. Such surveys, with millions of observations over half of the sky, will allow to identify the missing faint dwarf galaxy population, allowing for detailed studies of the AGN activity in dwarf galaxies.

\section{Summary}\label{sec:summary}
We select a sample of 1,058 VIPERS AGN in low-mass galaxies (log ${\rm M_\mathrm{*}/M}_{\odot})\lesssim10$) based on the emission line diagnostic diagram observed at redshift $0.5<\rm{z}<0.9$. 
Comparison to a SF(control) sample of 1,058 star-forming galaxies matched in stellar mass, ri colour and redshift, indicates that the AGN and non-AGN dwarf galaxies similarly reside in LD environments. The environment is quantified as the local density contrast derived from the fifth nearest neighbour technique. In particular, we find that:
\begin{itemize}
    \item Dwarves are preferably found in LD environments: 52\% AGN and 49\% SF(control) galaxies are found in LD environments (corresponding to voids) with only 10\% AGN and 10\% SF(control) galaxies located in HD environments (corresponding to galaxy groups and clusters; see Tab.~\ref{table:number_in_HD_LD}). 
    \item Dwarf galaxies are highly star-forming galaxies, located on the MS and starburst region (see Fig.~\ref{fig:MS}). 
    \item There are no visible differences in the $\delta$ distributions of AGN and SF(control) samples (see Fig.~\ref{fig:delta_distribution}); however, their similarity is confirmed only in one redshift bin ($\rm{0.65<z\leq0.80}$) with KS test and skewness values (see Tab.~\ref{tables:skewness}). 
    \item There is no sign of altering dwarf galaxy properties (SFR, D4000, $R_{\mathrm{e}}$) whether hosting AGN or not or moving to denser environments. 
    \item The SFR(OII) increase with AGN power is observed for all dwarf galaxies hosting or non-hosting AGN and is also independent of the environment (with significance of $<2\sigma$, see Fig.~\ref{fig:MS}).  
    \item AGN host over-massive black holes (with a median of log $\rm{M_{BH}}$ = 8.2 M$_{\odot}$) independently of the environment in which they reside. 
\end{itemize}

What turns on AGN activity in dwarf galaxies and how AGN impact the host properties is still unclear due to complex physical mechanisms regulating galaxy evolution as well as to limited observations of dwarf galaxies beyond the local Universe. 
For the first time, we show that dwarf galaxies hosting or not AGN reside in similar environments at so far unexplored redshift range of $\rm{0.4<z<0.9}$. 
Thanks to larger statistical sample we are able to probe a redshift evolution that we find none. 
In line with other studies at z$<0.5$~\citep[e.g.][]{Kristensen2020,Davis2022} our findings suggest that the environment played a negligible role (if any) in triggering AGN in low-mass galaxies already since $\rm{z=0.9}$. 
Furthermore, our findings show that AGN do not alter properties of the host suggesting that AGN are not gas-enriched due to environmental effects and do not show evidence for suppressing star formation at least since $\rm{z=0.9}$.

\section*{Acknowledgements}
The authors want to thank Thibaud Moutard, Giustina Vietri, Olga Cucciati, and Bianca Garilli for sharing with stellar mass, local density and line measurements for VIPERS galaxies. 
This work has been supported by the Polish National Agency for Academic Exchange (Bekker grant BPN/BEK/2021/1/00298/DEC/1), the European Union's Horizon 2020 Research and Innovation Programme under the Maria Sklodowska-Curie grant agreement (No. 754510) and the Spanish Ministry of Science and Innovation through the Juan de la Cierva-formacion programme (FJC2018-038792-I). MM acknowledges support from the Ramon y Cajal fellowship (RYC2019-027670-I). This work was also partially supported by the program Unidad de Excelencia Mar\'ia de Maeztu CEX2020-001058-M.

\section*{Data Availability}
The data underlying this article comes from VIPERS and the spectra can be accessed at the survey website: \url{http://vipers.inaf.it/}. The samples of AGN and SF(control) with physical and morphological parameters and local densities used in this paper will be released with the published version of the paper.



\bibliographystyle{mnras}
\bibliography{vipers}




\appendix
\section{Sanity checks}~\label{app:sanity_check}
In this Section, we check which bias can be introduced to our findings by adapting different selection criteria to construct our samples. 
In particular, we consider the impact of: i) the stellar mass threshold to select dwarf galaxies (see Sec.~\ref{sec:sample_selection}), ii) the choice of the emission line diagnostic diagram (see Sec.~\ref{sec:BPT}), iii) the reliability of emission line measurements (see Sec.~\ref{sec:BPT}), iv) the addition of LINERS to our AGN sample (see Sec.~\ref{sec:BPT}), and v) the incorporation of WISE-based classification (see Sec.~\ref{sec:wise_selection}). 

As we already mentioned in Sec.~\ref{sec:sample_selection}, we slightly relax the criterion on the stellar mass by adopting a cut of $\rm{log}(M_\mathrm{*}/M_{\odot})\leq10$. 
Here, we verify if this selection may introduce biases to our analysis by creating a sample of AGN and SF(control) galaxies, hereafter AGN(dwarf), SF(dwarf), respectively, with a cut on stellar mass equal to the stellar mass of the LMC ($\rm{log}(M_\mathrm{*}/M_{\odot})=9.5$).   
We also validate the choice of the emission line diagnostic diagram proposed by~\cite{2010Lamareille} by selecting 599 AGN, hereafter AGN(MEx), from our AGN sample fulfilling the AGN criterion  proposed by \cite{Juneau2011,Juneau2014}. At the same time, the star-forming nature of SF(control) is confirmed for 90\% of the sample (949 galaxies, hereafter SF(MEx)) by their location on the MEx diagram. 
For VIPERS data, the use of reliable line measurements (satisfying conditions listed in Sec.~\ref{sec:BPT}) is recommended for scientific analysis~\citep{Vietri2022, Pistis2022}. 
To check if such cleaning of emission line measurements is sufficient, we distinguish a secure sample of AGN and star-forming dwarf galaxies, hereafter AGN(secure), and SF(secure), respectively, with an additional condition on the EW to be higher than 3$\sigma$ for the measurements of [OII]$\lambda3726$, H$\beta$, [OIII]$\lambda5007$ used to construct the emission line diagnostic diagram. 
We also check if including LINERs in our AGN sample may bias our results by creating a sample of AGN including only Seyferts and X-ray confirmed AGN, hereafter AGN(Seyfert). 
Finally, we consider the impact of adding 393 WISE-selected AGN into the AGN sample, hereafter AGN(WISE), as well as removing 42 of them which were identified among the SF(control) sample, hereafter, SF(WISE), see Sec.~\ref{sec:wise_selection} for details about mid-infrared AGN candidate selection.  

\begin{figure*}
	\centerline{\includegraphics[width=0.99\textwidth]{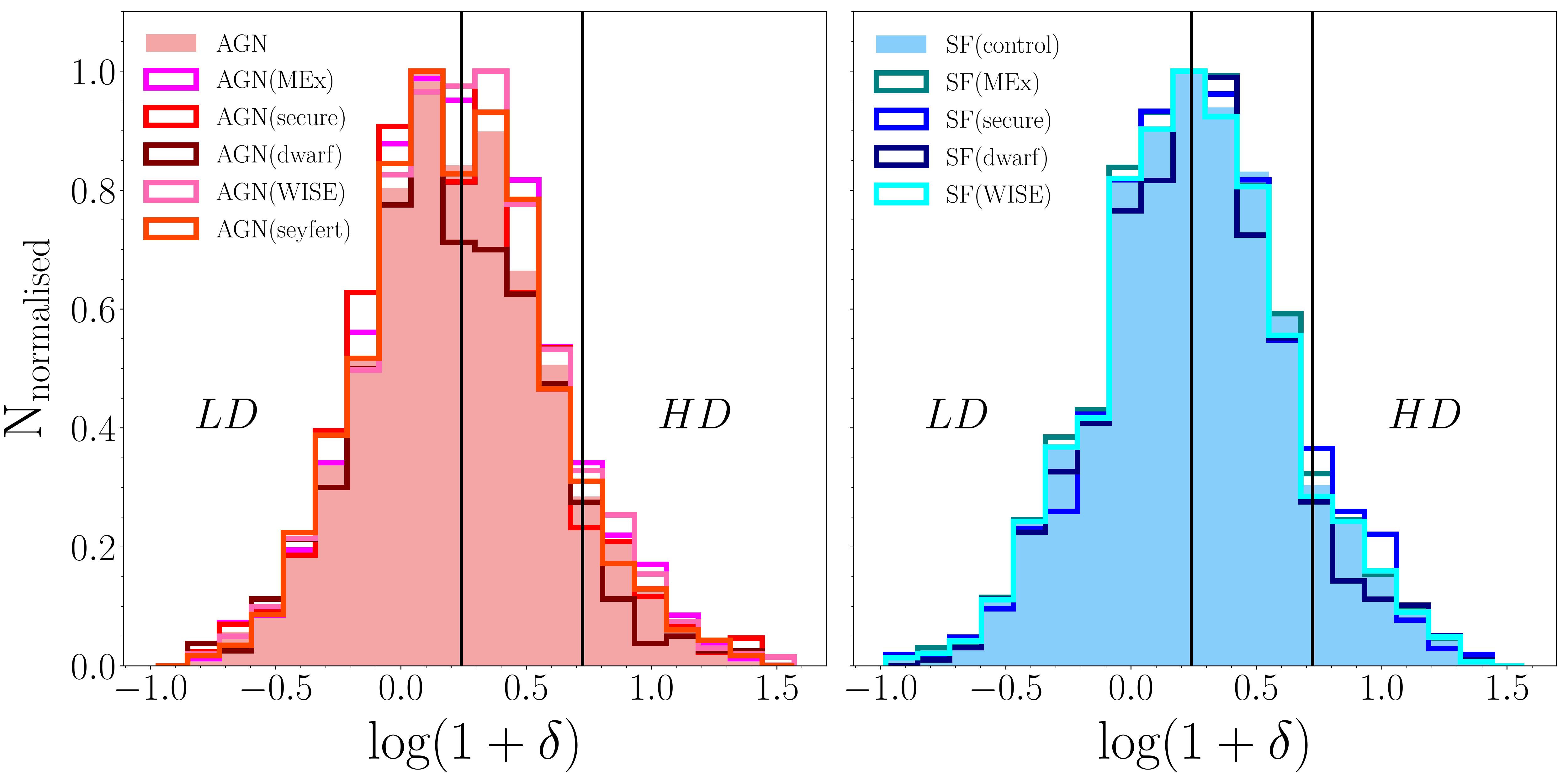}}
	\caption{Left panel: the normalised $\delta$ distribution for the AGN sample (filled histogram) and for the sanity samples: AGN(MEx) with AGN selected based on the MEx diagram,  AGN(secure) with secure emission line measurements, AGN(dwarf) with more restrictive cut on the stellar mass, AGN(WISE) with AGN selected based on the WISE diagram and AGN(Seyfert) with removing LINERS from the AGN sample. Right panel: the same but for the star-forming galaxy samples.   }
	\label{fig:density_distribution_app}
\end{figure*}

As shown in Fig.~\ref{fig:density_distribution_app}, the $\delta$ distributions for the AGN and SF(control) samples used in the paper have similar distribution as for the sanity samples: AGN(dwarf), AGN(MEx), AGN(secure), AGN(Seyfert), AGN(WISE), SF(dwarf), SF(MEx), SF(secure), and SF(WISE), as also confirmed by the KS test (see Tab.~\ref{tab:bias_app} and Appendix~\ref{app:statistical_tests} for more details about the KS test). 
The adopted criteria have also marginal effect on the percentage of AGN and star-forming galaxies residing in LD and HD environments, as shown in Table~\ref{tab:bias_app}. 
The only noticeable difference is in the percentage of AGN found in HD environments when using a more restrictive cut on stellar mass, when it drops from 10\% to 7\% compensated by the increase from 52\% to 55\% in LD environments. 
Nevertheless, this change for the AGN(dwarf) fraction does not affect significantly the fraction-$\delta$ relation, which shows slightly stronger negative trend but is still constant within $<2\sigma$. 
Similarly, for other AGN samples: for AGN(secure) and AGN(Seyfert) the trend of the relation is even stronger but consistent within $<0.5\sigma$. 
Interestingly, when incorporating the WISE-selected AGN candidates (AGN(WISE) sample) or selecting AGN based on the MEx diagram the trend of the relation is milder but still consistent with a significance of $1\sigma$. 
At the same time, removing the WISE-selected or MEx-selected AGN from the SF(control) sample (SF(WISE) and SF(MEx) samples) sharpen the fraction-$\delta$ relation, but it still is in agreement with the trend found for the SF(control) within $<0.2\sigma$. 
Also, the more restrictive criteria on selecting star-forming galaxies non-hosting AGN (SF(dwarf) sample) soften the negative trend of the fraction-$\delta$ relation found for the SF(control) sample, and the trends are in agreement within $<0.5\sigma$. 
Finally, selecting the SF(secure) sample reverses the trend of the fraction-$\delta$ relation found for the SF(control) sample, but the trends are in agreement within $<0.5\sigma$. 
Our sanity checks confirm the reliability of our findings, adapting more restrictive criteria on stellar mass cut, quality of emission line measurements or AGN sample composition does not alter our results by more than $<0.5\sigma$. 
This suggests that AGN selection based on the MEx diagram or the mid-infrared colours or defining dwarf galaxies as galaxies with $\rm{log(M_{*}/M_{\odot})<10}$  still leads to the same conclusion of no difference between environments in which dwarf galaxies hosting and non-hosting AGN reside.  

	\begin{table}
		\centering                         
		\begin{tabular}{P{1.5cm} P{1.cm} P{1.cm} P{1.0cm} P{1.0cm}}    
			\hline 
			sample  & N & $\rm{\%_{LD}}$ & $\rm{\%_{HD}}$ & $\rm{p_{KS}}$\\
			\hline 
            \hline
			AGN & 1,058 & 52 & 10 &   \\
            \hline
			AGN(dwarf) & 480 & 55 & 7 & 0.59  \\
			AGN(MEx) & 599 & {50} & {11} & {0.95}  \\
			AGN(secure) & 300 & 53 & 9  & 0.76\\
			AGN(Seyfert) & 795 & 52 & 10 & 1.00 \\
			AGN(WISE) & 1,451 & 50 & 10 & 0.80 \\
			\hline
			SF(control) & 1,058 & 49 & 10 &   \\
            \hline
			SF(dwarf) & 652 & 49 & 9 & 0.95  \\
			{SF(MEx)} & {949} & {49} & {10} & {0.84}  \\
			SF(secure) & 743 & 48 & 11  & 0.97\\
			SF(WISE) & 1,016 & 50 & 10  & 0.80\\
			
			\end{tabular}
		\caption{Number and percentage in LD and HD environments for each sample of AGN and star-forming dwarf galaxies. 
			 }             
		\label{tab:bias_app}     
	\end{table}

\begin{figure*}
	\centerline{\includegraphics[width=0.99\textwidth]{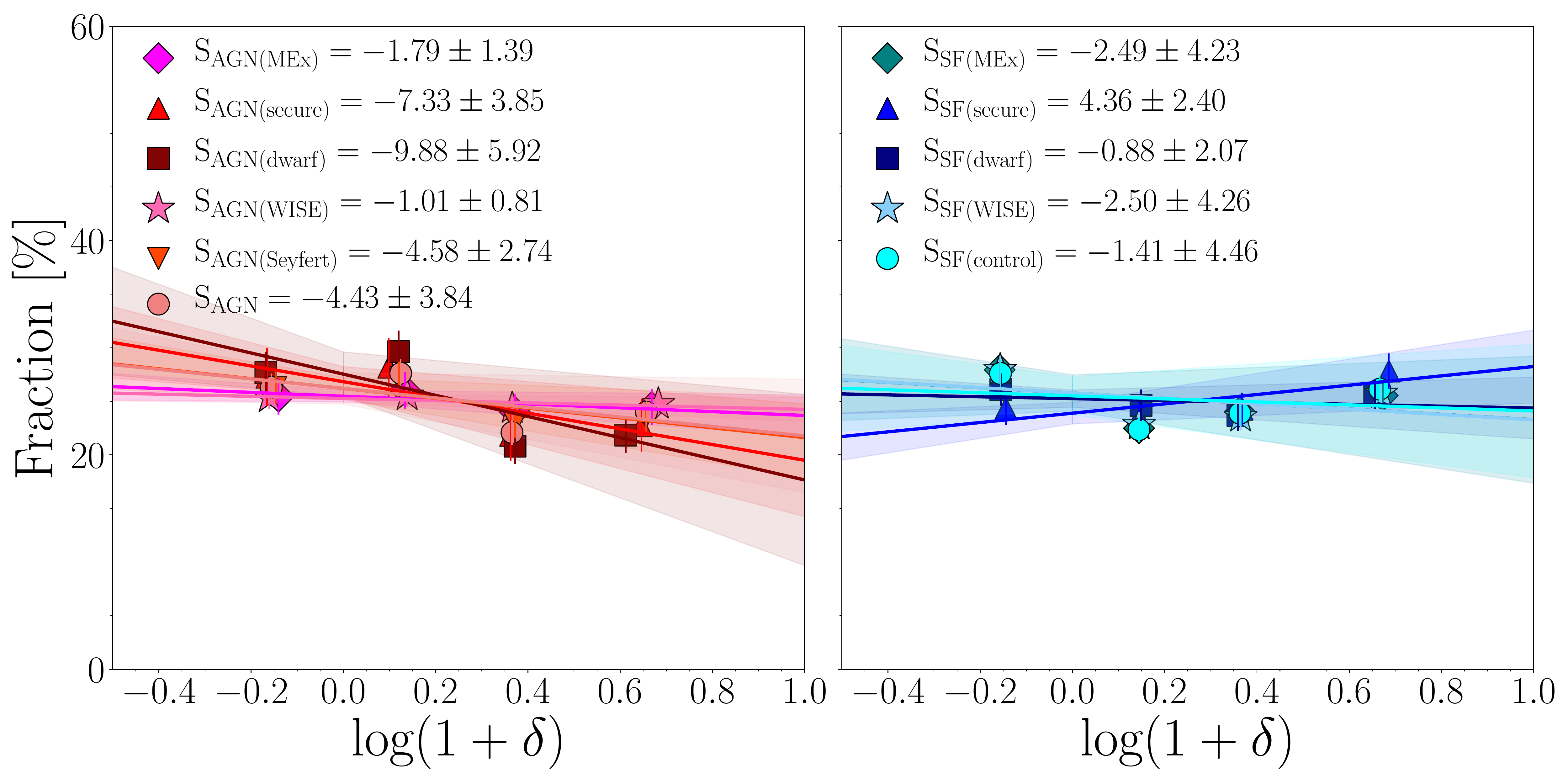}}
	\caption{Left panel: fraction-$\delta$ relation for the AGN sample and for the sanity samples: AGN(MEx) with AGN selected based on the MEx diagram, AGN(secure) with secure emission line measurements, AGN(dwarf) with more restrictive cut on the stellar mass, AGN(WISE) with AGN selected based on the WISE diagram, and AGN(Seyfert) with including solely Seyfert 2 galaxies in the AGN sample. Right panel: the same but for the star-forming samples.   }
	\label{fig:density_fraction_app}
\end{figure*} 

\section{Properties the of SF(control) sample}\label{app:statistical_tests}

\begin{figure*}
	\centerline{\includegraphics[width=0.99\textwidth]{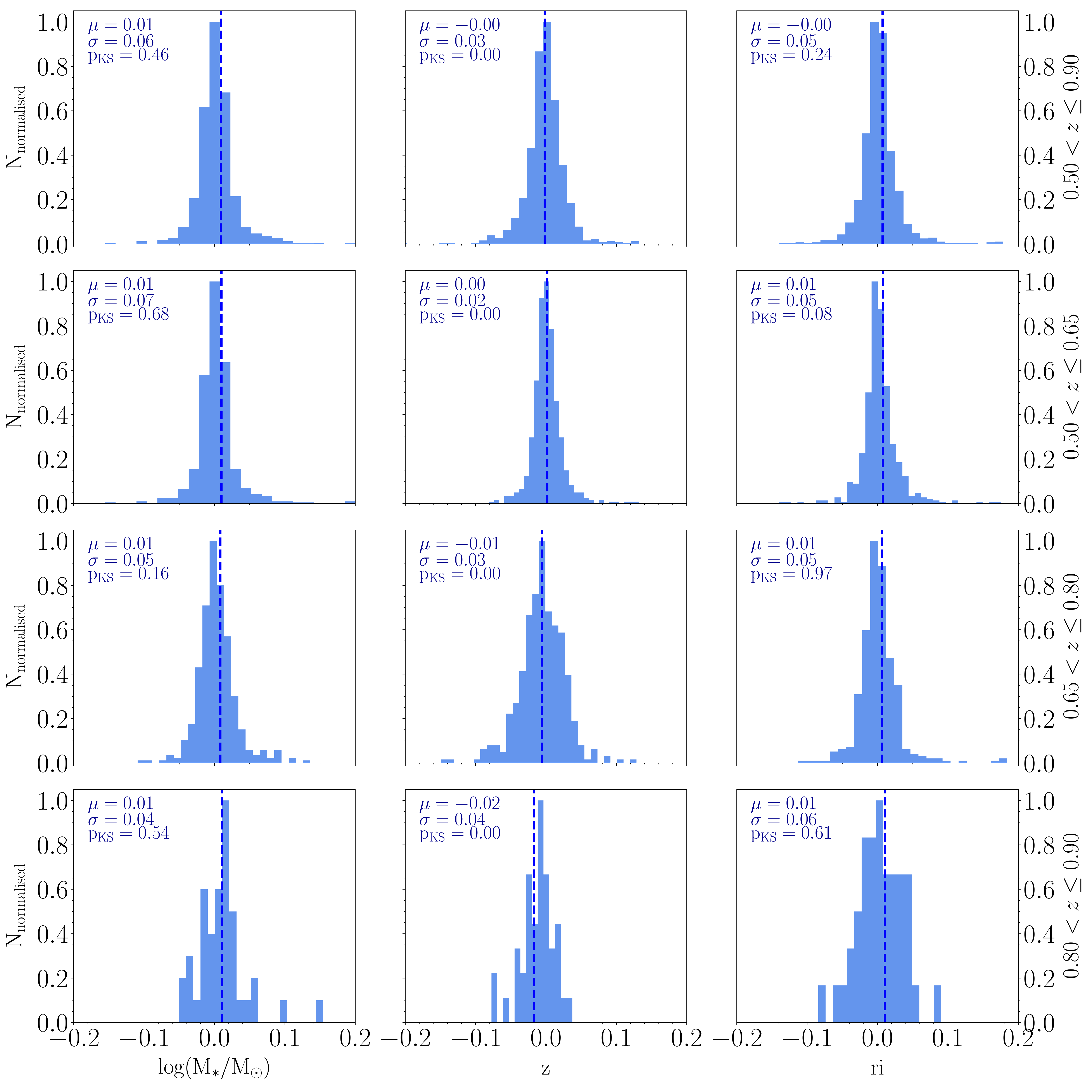}}
	\caption{Distributions of the difference between stellar mass, redshift, and $\rm{ri}$ colour of the AGN and SF(control) samples are given in the left, middle, and right panels, respectively. The differences found for the whole redshift range are shown in the upper row, while for the three different redshift bins in the remaining rows. }
	\label{fig:param_dif}
\end{figure*}

To demonstrate the quality of the SF(control) sample in recreating similar range of parameters as the AGN sample, Fig.~\ref{fig:param_dif} presents the distribution of  differences in physical parameters between galaxies in the AGN and SF(control) samples.  
It is clear that the distributions of their main properties are similar, showing a mean difference on the level of $0.01\pm0.06$,$0.00\pm0.03$,$0.00\pm0.05$ for stellar mass, redshift and $\rm{ri}$ colour, respectively. 
This tendency is also preserved in each redshift bin. 

To check if the small differences between the SF(control) and AGN samples (see Fig.~\ref{fig:param_dif} and Table~\ref{table:properties}) are statistically significant, we calculate the statistical tests:  Kolmogorov-Smirnov~\citep[KS;][]{Massey1951}, Anderson-Darling~\citep[AD;][]{Anderson1952} and Mann-Whitney-Wilcoxon~\citep[MWW;][]{Mann1947}.  
The KS test is commonly used non-parametric test  based on the distance between two cumulative distributions. 
The AD test, similarly to the KS test, is based on the empirical distribution function (EDF) but more sensitive to differences in tails of the distributions of samples in question~\citep{Engmann2013}. 
Motivated by the small sample size, we also calculate MWW test, which is more sensitive to differences than the KS test when sample sizes are small~\citep{Suphi2013}.
We test the null probability for each pair of AGN-SF(control) parameter distributions and the results for each test are given in Table~\ref{table:KStest}. 
Low p-values indicate that the two distributions in question are probably not from the same underlying distribution, meaning is highly probable that the two distributions are different. 
Then, high p-values indicate that the two samples are statistically indistinguishable or more accurately we cannot reject the hypothesis that the distributions of the two samples are the same. 

Under all of these tests we cannot reject the hypothesis that the AGN and SF(control) distributions of masses and colours are drawn from the same population. 
This suggests that the AGN sample has the stellar mass and $\rm{ri}$ colour indistinguishable from the SF(control) sample for all redshift bins considered in this work. 
The redshift distributions are also different as pointed by low p-values (except for AD and MWWW tests at the full redshift range).   
Similarly,~\citealt{Cheung2015} found that the redshift distribution is not consistent ($\rm{p_{KS}=0.03}$) for the AGN and control samples drawn from 1,023 GOOD-S galaxies observed within the redshift range $0.2<z<1.0$. 
However,~\citealt{Cheung2015} showed that their results from the GOODS-S sample are consistent with those drawn from other sets characterised by similar redshift distribution of the AGN and control samples (with $\rm{p_{KS}=0.94(0.50)}$ for the AEGIS(COSMOS) sample) indicating that the low p-values for the redshift distributions of the AGN-control samples does not bias the results. 
Therefore, we perform the analysis in three different redshift bins, but when interpreting our results we have to keep these facts in mind and the main conclusion should be rather drawn when investigating the whole redshift range. 


    \renewcommand{\arraystretch}{1.4}

	\begin{table*}
		\centering                         
		\begin{tabular}{p{2.0cm}  p{1.0cm} p{1.0cm} p{1.2cm}  p{1.0cm} p{1.0cm} p{1.2cm}  p{1.0cm} p{1.0cm} p{1.0cm}}   
		    \cline{2-10}
			 \multirow{2}{*}{}& \multicolumn{3}{c}{$\rm{M_{*}}$} &  \multicolumn{3}{c}{$z$} & \multicolumn{3}{c}{$ri$} \\
		    \cline{2-10}
			 & $p_{KS}$ &  $p_{AD}$ & $p_{MWW}$ & $p_{KS}$ &  $p_{AD}$ & $p_{MWW}$ & $p_{KS}$ &  $p_{AD}$ & $p_{MWW}$\\			
			\hline 
			0.50$<$z$\leq$0.90 & 0.46 & $\geq$0.25 & 0.16 & \textbf{$\leq$0.01} & 0.07 & 0.29 & 0.47 & 0.16 & 0.05  \\
			\hline
 			0.50$<$z$\leq$0.65 & 0.68 & 0.23 & 0.09 &\textbf{$\leq$0.01} & \textbf{$\leq$0.01} & \textbf{$\leq$0.01} & 0.08 & 0.03 & 0.02  \\
 			0.65$<$z$\leq$0.80 & 0.16 & 0.09 & 0.06 & \textbf{$\leq$0.01} & \textbf{$\leq$0.01} & \textbf{$\leq$0.01} & 0.97 & $\geq$0.25 & 0.43  \\
 			0.80$<$z$\leq$0.90 & 0.54 & $\geq$0.25 & 0.15 & \textbf{$\leq$0.01} & \textbf{$\leq$0.01} & \textbf{$\leq$0.01} & 0.61 & $\geq$0.25 & 0.16  \\ 			

		\end{tabular}
		\caption{Statistical test of differences between the distributions of properties of the SF(control) and AGN samples. The p-values of Kolmogorov-Smirnov ($\rm{p_{KS}}$), Anderson-Darling ($\rm{p_{AD}}$) and Mann-Whitney-Wilcoxon ($\rm{p_{MWW}}$) tests return the probabilities under the null hypothesis that the two samples are drawn from the same population. The probabilities indicating that the null hypothesis is rejected (i.e the distribution are \textit{not} drawn from the same population) have low values ($p\leq$0.01). 
		The p-values above the significance level (p-values$\leq0.01$) indicate that there are no statistically significant differences in the distributions of two samples. The probabilities with significant level below to our threshold are shown in bold.  
			 }             
		\label{table:KStest}     
	\end{table*}


\bsp	
\label{lastpage}
\end{document}